\documentclass[10pt,draftcls, onecolumn]{IEEEtran}
\usepackage{amsmath,amssymb,mathrsfs}
\usepackage{epsfig,epsf,subfigure,graphicx,graphics,latexsym}
\usepackage{cite, url, hyperref}
\usepackage[usenames,dvipsnames]{pstricks}
\usepackage{pst-plot}
\usepackage{pst-node}
\newtheorem{lemma}{Lemma}
\newcommand{\lp}{\left(}
\newcommand{\rp}{\right)}
\newcommand{\Ppr}{\frac{1}{a^2}}
\newcommand{\C}{\mathcal{C}}
\begin{document}
\title{The Approximate Capacity Region of the Gaussian Z-Interference Channel with Conferencing Encoders}
\author{\authorblockN{\small Hossein Bagheri, Abolfazl
S. Motahari, and Amir K. Khandani}\\
\authorblockA{\small Department of Electrical and Computer Engineering, University of Waterloo\\
Emails: \{hbagheri, abolfazl,
khandani\}@cst.uwaterloo.ca}
\thanks{Financial support provided by Nortel and the
corresponding matching funds by the Natural Sciences and Engineering
Research Council of Canada (NSERC), and Ontario Ministry of Research
\& Innovation (ORF-RE) are gratefully acknowledged.}}
\maketitle
\begin{abstract}
A two-user Gaussian Z-Interference Channel (GZIC) is considered, 
in which encoders are connected through noiseless links with finite capacities. In this setting, prior to each transmission block the encoders communicate with
each other over the cooperative links. The capacity region and the sum-capacity of the channel are characterized
within 1.71 bits per user and 2 bits in total, respectively. It is also established that properly sharing the total limited cooperation capacity between the cooperative links may enhance the achievable region, even when compared to the case of unidirectional transmitter cooperation with infinite cooperation capacity. To obtain the results, genie-aided upper bounds on the sum-capacity and cut-set bounds on the individual rates are compared with the achievable rate region. In the interference-limited regime, the achievable scheme enjoys a simple type of Han-Kobayashi signaling, together with the zero-forcing, and basic relaying techniques. In the noise-limited regime, it is shown that treating interference as noise achieves the capacity region up to a single bit per user.
 \end{abstract}
\section{Introduction}
Interference limits the throughput of a network, consisting of multiple non-cooperative transmitters, intending to convey independent messages to their corresponding receivers through a common bandwidth. The way interference is usually dealt with is by either treating it as noise or preventing it by associating different orthogonal dimensions, e.g. time or frequency division to different users. Since interference has structure, it is possible for a receiver to decode some part of the interference and remove it from the received signal. This is indeed the coding scheme proposed by Han-Kobayashi (HK) for the two-user Gaussian Interference Channel (GIC) \cite{HKIT81}. The two-user GIC provides a simple example showing that a single strategy against interference is not optimal. In fact, one needs to adjust the strategy according to the channel parameters \cite{Abolfazl_ISIT_08, Shang_ISIT_08, Annapureddy_ISIT_08}. However, a single suboptimal strategy can be proposed to achieve up to 1 bit per user of the capacity region of the two-user GIC \cite{EtkinIT07}.

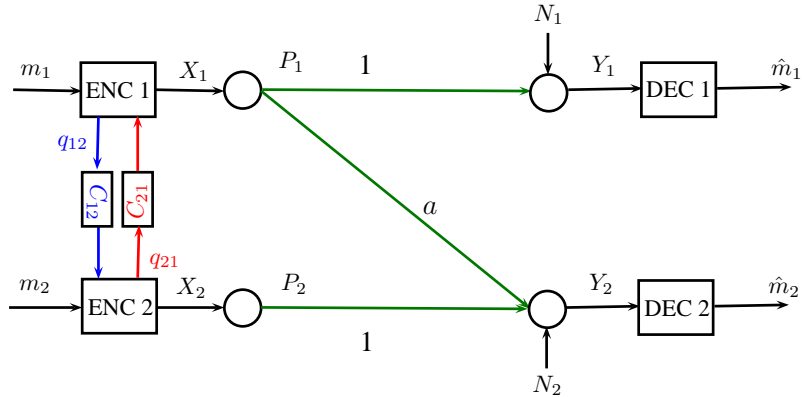
\begin{figure}
\begin{center}
\scalebox{.9}
{
\begin{pspicture}(0,-2.948125)(12.562813,2.948125)
\definecolor{color1410}{rgb}{0.0,0.47058823529411764,0.0}
\pscircle[linewidth=0.04,dimen=outer](3.5321877,1.6428127){0.2912501}
\psline[linewidth=0.04cm,linecolor=color1410,arrowsize=0.05291667cm 2.0,arrowlength=1.4,arrowinset=0.4]{->}(3.8234375,1.6515627)(7.8009377,1.6296875)
\psline[linewidth=0.04cm,linecolor=color1410,arrowsize=0.05291667cm 2.0,arrowlength=1.4,arrowinset=0.4]{->}(3.8034377,1.6315626)(7.7809377,-1.5903125)
\usefont{T1}{ptm}{m}{n}
\rput{-90.0}(1.3876561,1.3214065){\rput(1.3354688,-0.01343735){\color{blue}$C_{12}$}}
\usefont{T1}{ptm}{m}{n}
\rput(5.347031,2.0315626){\large 1}
\usefont{T1}{ptm}{m}{n}
\rput(5.367031,-2.0684373){\large 1}
\usefont{T1}{ptm}{m}{n}
\rput(6.3014064,-0.08843736){\large $a$}
\usefont{T1}{ptm}{m}{n}
\rput{-270.0}(2.0214067,-1.9676563){\rput(1.975469,0.04656265){\color{red}$C_{21}$}}
\pscircle[linewidth=0.04,dimen=outer](8.032187,-1.5971873){0.2912501}
\pscircle[linewidth=0.04,dimen=outer](8.052188,1.6028126){0.2912501}
\usefont{T1}{ptm}{m}{n}
\rput(2.8123438,1.9196875){$X_1$}
\psframe[linewidth=0.04,dimen=outer](2.2609375,2.0696876)(1.1209375,1.2296875)
\usefont{T1}{ptm}{m}{n}
\rput(1.6951562,1.6396875){ENC 1}
\psframe[linewidth=0.04,dimen=outer](2.2809374,-1.1303124)(1.1409374,-1.9703125)
\usefont{T1}{ptm}{m}{n}
\rput(1.7296875,-1.5603125){ENC 2}
\usefont{T1}{ptm}{m}{n}
\rput(2.7723436,-1.2803125){$X_2$}
\usefont{T1}{ptm}{m}{n}
\rput(0.48234376,1.9196875){$m_1$}
\psline[linewidth=0.04cm,arrowsize=0.05291667cm 2.0,arrowlength=1.4,arrowinset=0.4]{->}(0.0809375,-1.5703125)(1.1609375,-1.5703125)
\usefont{T1}{ptm}{m}{n}
\rput(0.46234375,-1.2603126){$m_2$}
\psline[linewidth=0.04cm,arrowsize=0.05291667cm 2.0,arrowlength=1.4,arrowinset=0.4]{->}(8.340938,1.6496875)(9.440937,1.6696875)
\usefont{T1}{ptm}{m}{n}
\rput(8.872344,1.9996876){$Y_1$}
\psframe[linewidth=0.04,dimen=outer](10.540937,2.0696876)(9.400937,1.2296875)
\usefont{T1}{ptm}{m}{n}
\rput(11.572344,1.9596875){$\hat{m}_1$}
\psline[linewidth=0.04cm,arrowsize=0.05291667cm 2.0,arrowlength=1.4,arrowinset=0.4]{->}(10.540937,1.6696875)(11.640938,1.6896875)
\psline[linewidth=0.04cm,arrowsize=0.05291667cm 2.0,arrowlength=1.4,arrowinset=0.4]{->}(8.300938,-1.5703125)(9.400937,-1.5503125)
\usefont{T1}{ptm}{m}{n}
\rput(8.832344,-1.2203125){$Y_2$}
\psframe[linewidth=0.04,dimen=outer](10.500937,-1.1503125)(9.360937,-1.9903125)
\usefont{T1}{ptm}{m}{n}
\rput(11.532344,-1.2603126){$\hat{m}_2$}
\psline[linewidth=0.04cm,arrowsize=0.05291667cm 2.0,arrowlength=1.4,arrowinset=0.4]{->}(10.500937,-1.5503125)(11.600938,-1.5303125)
\usefont{T1}{ptm}{m}{n}
\rput(2.3823438,-0.8803125){\color{red}$q_{21}$}
\usefont{T1}{ptm}{m}{n}
\rput(1.0223438,0.8596875){\color{blue}$q_{12}$}
\usefont{T1}{ptm}{m}{n}
\rput(9.975781,1.6596875){DEC 1}
\usefont{T1}{ptm}{m}{n}
\rput(9.950313,-1.5803125){DEC 2}
\psline[linewidth=0.04cm,arrowsize=0.05291667cm 2.0,arrowlength=1.4,arrowinset=0.4]{->}(2.2609375,-1.5703125)(3.2609375,-1.5703125)
\pscircle[linewidth=0.04,dimen=outer](3.5321877,-1.5771873){0.2912501}
\psline[linewidth=0.04cm,linecolor=color1410,arrowsize=0.05291667cm 2.0,arrowlength=1.4,arrowinset=0.4]{->}(3.8034375,-1.5684373)(7.7409377,-1.5903125)
\usefont{T1}{ptm}{m}{n}
\rput(4.2923436,-1.2203125){$P_2$}
\usefont{T1}{ptm}{m}{n}
\rput(4.2523437,2.0596876){$P_1$}
\psframe[linewidth=0.04,dimen=outer](2.2209375,0.4496875)(1.7409375,-0.3903125)
\psline[linewidth=0.04cm,linecolor=red,arrowsize=0.05291667cm 2.0,arrowlength=1.4,arrowinset=0.4]{->}(1.9809375,0.4496875)(1.9809375,1.2696875)
\psline[linewidth=0.04cm,linecolor=red,arrowsize=0.05291667cm 2.0,arrowlength=1.4,arrowinset=0.4]{->}(1.9809375,-1.1303124)(2.0009375,-0.3503125)
\psline[linewidth=0.04cm,linecolor=blue,arrowsize=0.05291667cm 2.0,arrowlength=1.4,arrowinset=0.4]{->}(1.4009376,1.2496876)(1.3809375,0.4696875)
\psline[linewidth=0.04cm,linecolor=blue,arrowsize=0.05291667cm 2.0,arrowlength=1.4,arrowinset=0.4]{->}(1.4009376,-0.3503125)(1.4009376,-1.1503125)
\psframe[linewidth=0.04,dimen=outer](1.6209375,0.4496875)(1.1409374,-0.3903125)
\psline[linewidth=0.04cm,arrowsize=0.05291667cm 2.0,arrowlength=1.4,arrowinset=0.4]{->}(2.2409375,1.6496875)(3.2609375,1.6296875)
\psline[linewidth=0.04cm,arrowsize=0.05291667cm 2.0,arrowlength=1.4,arrowinset=0.4]{->}(0.1409375,1.6496875)(1.1609375,1.6296875)
\usefont{T1}{ptm}{m}{n}
\rput(8.052343,2.7596874){$N_1$}
\usefont{T1}{ptm}{m}{n}
\rput(8.032344,-2.7203126){$N_2$}
\psline[linewidth=0.04cm,arrowsize=0.05291667cm 2.0,arrowlength=1.4,arrowinset=0.4]{->}(8.020938,-2.4703126)(8.020938,-1.8503125)
\psline[linewidth=0.04cm,arrowsize=0.05291667cm 2.0,arrowlength=1.4,arrowinset=0.4]{->}(8.040937,2.4896874)(8.040937,1.8696876)
\end{pspicture}
}
\caption{The Z-Interference Channel with Conferencing Encoders. The cooperation links are orthogonal to each other and to the communication medium as shown with different colors in the figure. In the proposed achievable schemes, the cooperative links with capacities $C_{12}$ and $C_{21}$ are used for zero-forcing and relaying purposes, respectively.}\label{fig:model}
\end{center}
\end{figure}

If the senders can cooperate, interference management can be done more effectively through cooperation. Cooperative links can be either orthogonal or non-orthogonal to the shared medium. In this work, orthogonal cooperation is considered. In addition, in order to understand some fundamental aspects of the optimal coding scheme (in the sense of having a constant gap to the capacity region), the GZIC is investigated, in which one transmitter-receiver pair is interference-free (see Fig. \ref{fig:model}.).

\textbf{\emph{Prior Works}}. Transmitter coordination over
orthogonal links is studied for different scenarios with
two transmitters (cf. \cite{WillemsIT83, NgIT071,
MaricIT07, DevroyeIT06, WuIT07, JovicicIT06, Rini_ITW10, Hossein_GICUC, Hossein_GICBC}). The capacity region of the Multiple Access Channel (MAC)
with cooperating encoders is derived in \cite{WillemsIT83},
where cooperation is referred to as \emph{conference}. Several achievable rate
regions are proposed for the GIC with bidirectional transmitter and receiver
cooperation \cite{NgIT071}. In the
transmit cooperation, the entire message of each transmitter is
decoded by the other cooperating transmitter, which apparently
limits the performance of the scheme to the capacity of the involved
cooperative link. The capacity regions of the compound MAC with
conferencing encoders and the GIC with degraded message set, under certain strong interference conditions, are obtained in \cite{MaricIT07}. The GIC with degraded message set is also termed \emph{Gaussian Cognitive Radio (GCR)} in the literature \cite{DevroyeIT06}. The GCR can be considered as a GIC with unidirectional orthogonal cooperation, in which the capacity of the cooperative link is infinity.\footnote{Technically, the capacity of the cooperative link needs to be equal to the message rate of the user sending data via the cooperative link \cite{WillemsIT83}.} The capacity region of the GCR is established for the weak interference regime in \cite{WuIT07, JovicicIT06}. Recently, the capacity region of the GCR is characterized within 1.87 bits for all ranges of the channel parameters\cite{Rini_ITW10}.
Furthermore, the sum-capacity of the symmetric GIC with unidirectional and bidirectional cooperation is approximated up to 2 and 2.13 bits, respectively in \cite{Hossein_GICUC, Hossein_GICBC}. The achievable schemes are based on a simple type of
HK scheme, cooperative communication, and zero-forcing technique.
For non-orthogonal cooperative links, an achievable rate region is proposed in \cite{Chao_ISIT_07} and the sum-capacity of the GIC is determined up to 18 bits \cite{Prabhakaran09}. Very recently, in a parallel and independent work, the capacity region of the GIC with bidirectional cooperation is characterized within 6.5 bits\cite{Tse_cof_EncIT10}.\footnote{An earlier version of our work containing most of the results is reported in Library and Archives Canada Technical Report UW-ECE 2010-04, Feb. 2010 \cite{Hossein_Tech_ZBC}.}

\textbf{\emph{Contributions and Relation to Previous Works}}.
Simple achievable schemes based on HK, relaying and zero-forcing techniques are shown to achieve the capacity region within 1.71 bits per user for \emph{all} channel parameters.\footnote{It is remarked that the binning technique \cite{GelfandPr80} can be used at the encoder corresponding to the receiver with interference, to precode its message against the known interference. This, in general, could enlarge the achievable rate region. However, it is shown that one can achieve the capacity region within 1.71 bits without using the binning technique.} In this paper, some important features of the problem are explored one-by-one and the appropriate achievable schemes are proposed accordingly. In the first step, the GZIC with unidirectional cooperation is considered. It is demonstrated that the HK scheme together with zero-forcing or relaying can achieve the capacity region up to 1.5 bits per user. Then, based on the observations made in the unidirectional case, the capacity region of the GZIC with bidirectional cooperation is determined up to 1.71 bits per user. Our step-by-step approach to solve the problem is in contrast to the universal strategy of \cite{Tse_cof_EncIT10}, in which the same signaling is used for all channel parameters. Applying the scheme of \cite{Tse_cof_EncIT10} to the GZIC, five signals should be jointly decoded at each receiver, whereas three signals are required to be jointly decoded in our paper, which simplifies the transmission scheme. Appropriate compression and sequential decoding techniques are utilized to facilitate such a low complexity decoding. In the achievable schemes, proper power allocation over the employed codewords plays an essential role to achieve the result. The Linear Deterministic Model (LDM) proposed in \cite {IT:Avestimehr} is incorporated to attain such a power allocation. It is illustrated that for some channel parameters, no cooperation or unidirectional cooperation is sufficient to obtain the results. It is also argued that a suitable distribution of total cooperation capacity between the cooperative links can enhance the rate region. In particular, it is demonstrated that the achievable region of the GZIC with limited bidirectional cooperation may outperform the capacity region of the GZIC with infinite unidirectional cooperation, known as the cognitive Z channel.
When the noise is the performance-limiting factor instead of the interference, it is shown that treating interference as noise, and not using the cooperative links, achieve within 1 bit of the capacity region of the channel for all channel parameters.

The rest of the paper is structured as follows. Section \ref{sec: sys_model} describes the system model and the preliminaries. Section \ref{sec: Upper Bound} presents an outer bound on the capacity region. Section \ref{sec: UC} focuses on the unidirectional cooperation case and provides achievable schemes employing zero-forcing or relaying techniques depending on which cooperative link is present. Then, Section \ref{sec: BC} adjusts the achievable schemes proposed for the unidirectional case to the general bidirectional cooperation case. Section \ref{sec: BCvsUC} numerically compares the performance of bidirectional and unidirectional cooperation cases. Finally, Section \ref{sec: Conclusion} concludes the paper. The detailed proofs and gap analysis are left to the appendices.

\textbf{\emph{Notation}}.
Throughout the paper, all logarithms are to base 2, and $\mathcal{C}(P)\!\triangleq\!\frac{1}{2}\log \left(1+P\right)$. The set of $\epsilon$ jointly typical sequences of random variables $(x^n, y^n)$, with length $n$, is represented by $A_{\epsilon}^{(n)}(x^n, y^n)$ (for the definition, see \cite{book:Cover}). Whenever it is clear from the context, the random variables may be omitted from the notation.
\section{System Model and Preliminaries}\label{sec: sys_model}
In this work, a two-user GZIC with partial transmit cooperation, as depicted in Fig. 1, is considered. The model consists of two transmitter-receiver pairs, in which each transmitter wishes to convey its own data to its corresponding receiver. There exist two noiseless
cooperative links with capacities
$C_{12}$ and $C_{21}$, respectively from Encoder $1$ to Encoder $2$ and vice versa. It is assumed that all nodes
are equipped with a single antenna. The input-output relationship for this channel in standard form is expressed as \cite{SasonIT04}:
\begin{equation}
\begin{array}{rl}
  Y_1\! &\!= X_1+N_1,\\
  Y_2\! &\!= aX_1+ X_2+N_2,
  \end{array}
\end{equation}
where $a\!\geq\!0$, and for $i\!\in\!\{1,2\}$, $N_{i}\!\sim\!{\cal N}(0, 1)$, \emph{i.e.}, is the Gaussian noise with zero mean and unit variance. The average
power constraint of the transmitters are respectively $P_1$ and $P_2$.
The full channel state information is assumed to be available at
both the transmitters and the receivers.

For a given block length $n$, Encoder $i\!\in\!\{1,2\}$
sends its own (random) message index $m_{i}$ from the index set
$\mathcal{M}_{i}\!=\!\{1,2,...,M_{i}\!=\!2^{nR_{i}}\}$ with rate $R_{i}$
[bits/channel use]. Each pair $(m_{1},m_{2})$ occurs with the same
probability $\frac{1}{M_{1}M_{2}}$. The result of the conference between the two encoders is two codewords $q_{12}^n, q_{21}^n$, where for $j\!\in\!\{1,2\}$, $i\!\neq\!j$, and each time index $t\!\in\!\{1,\cdots,n\}$, $q_{ij}[t]$ is only a function of $(m_i, q_{ji}[1], \cdots q_{ji}[t-1])$. The encoding function $f_{i}$ maps the
message index $m_{i}$ and $q_{ji}^n$ into a codeword $X_{i}^{n}$ chosen
from codebook $C_{i}$. Therefore:
\begin{equation}\label{eq: v_def}
\begin{array}{rl}
  X_{1}^{n} &= f_{1}(m_{1}, q_{21}^n),\\
  X_{2}^{n} &= f_{2}(m_{2},q_{12}^n).
  \end{array}
\end{equation}
The codewords in each codebook must satisfy the average power
constraint $\frac{1}{n}\sum\limits_{t=1}^{n}|X_{i}[t]|^2\!\leq\!P_i$.
 Each decoder uses a decoding function
$g_{i}(Y_{i}^{n})$ to decode its
desired message index $m_{i}$ based on its received sequence.
Let $\hat{m}_{i}$ be the output of the decoder. The
average probability of error for each decoder is
$P_{e_{i}}\!=\!\text{Pr}(\hat{m}_{i}\!\neq\!m_{i})$.
A rate pair ($R_{1}$, $R_{2}$) is said to be achievable when there
exists an ($M_{1},M_{2},n,P_{e_{1}},P_{e_{2}}$)-code for the GZIC
consisting of two encoding functions $\{f_{1}, f_{2}\}$ and two
decoding functions $\{g_{1}, g_{2}\}$ such that for sufficiently
large $n$:
\begin{equation*}
\begin{array}{rl}
  R_{1} &\leq \frac{1}{n} \log(M_{1}),\\
  R_{2} &\leq \frac{1}{n} \log(M_{2}), \\
  P_{e} &\leq \epsilon.
  \end{array}
\end{equation*}
In the above, $P_{e}\!=\!\max(P_{e_{1}}, P_{e_{2}})$ and $\epsilon\!>
\!0$ is a constant that can be chosen arbitrarily small. The capacity
region of the GZIC with conferencing encoders is the closure of the set of achievable rate pairs. The boundary of the achievable region ${\cal R}$ is said to be within $(\Delta_1,\Delta_2)$ of the boundary of the upper bound region ${\cal R}^{\text{up}}$ if for any pair $(R^{\text{up}}_1,R^{\text{up}}_2)$ on the boundary of the outer bound, there exists a pair $(R_1,R_2)$ on the achievable region such that $R^{\text{up}}_1-R_1\leq \Delta_1$, and $R^{\text{up}}_2-R_2\leq \Delta_2$. In this case, the achievable scheme leading to the region ${\cal R}$ is referred to as ${\cal R}(\Delta_1,\Delta_2)$ achievable.

If the regions associated with ${\cal R}^{\text{up}}$ and ${\cal R}(\Delta_1,\Delta_2)$ are polytopes, each facet of the achievable region is compared to its corresponding facet in the upper bound region. Defining $\delta_{R_1}\triangleq R^{\text{up}}_1-R_1$, we have $\delta_{R_1}\leq \Delta_1$. Similarly, for the facet related to $i R_1\!+\!j R_2$, for any $i, j\in\{1, 2, \cdots\}$, we have $\delta_{i R_1+j R_2}\leq i\Delta_1+ j\Delta_2$.

In this work, the interference-limited
regime is mainly investigated, \emph{i.e.}, $a^2 P_1\!\geq\!1$, since otherwise the system
is noise limited and is not of much interest. The noise-limited regime will be briefly considered in Section \ref{sec: BC}.
\section{Upper Bounds}\label{sec: Upper Bound}
\begin{lemma} The following region is an upper bound on the capacity region of the GZIC shown in Fig. \ref{fig:model}:
\begin{align}
R_1^{\text{up}}&\le \C(P_1)\label{eq: Ru1_1}\\
R_2^{\text{up}}&\le \C(P_2)+ C_{21}\label{eq: Ru2_1}\\
R_2^{\text{up}}&\le \C\left(2a^2P_1+2P_2\right)\label{eq: Ru2_2}\\
R_1^{\text{up}}\!+\!R_2^{\text{up}}\!&\!\le \C(\frac{\max\{1\!-\!a^2,0\}P_1}{1\!+\!a^2P_1})\!+\!\C(2a^2P_1\!+\!2 P_2)\!+\!C_{12}\label{eq: Ru1pRu2_1}\\
R_1^{\text{up}}+R_2^{\text{up}} &\le \C\lp P_1P_2+P_1(1+2 a^2)+2 P_2 \rp.\label{eq: Ru1pRu2_2}
\end{align}
\end{lemma}
\begin{IEEEproof}
The first three bounds are simple applications of cut-set bounds at Transmitter 1, Transmitter 2, and Receiver 2, respectively. The sum-rate upper bounds are the tailored version of the bounds obtained for the GIC with conferencing encoders in \cite{Hossein_GICBC}. To simplify the gap analysis, the term $(\sqrt{a^2P_1}+\sqrt{P_2})^2$ in \cite{Hossein_GICBC} is replaced by the larger term $2a^2P_1+2P_2$ in (\ref{eq: Ru2_2}), (\ref{eq: Ru1pRu2_1}), and (\ref{eq: Ru1pRu2_2}).
\end{IEEEproof}
\section{Unidirectional Cooperation}\label{sec: UC}
To gain some insight into the essential ingredients of an appropriate achievable scheme, first, it is assumed that one of the cooperative links has zero capacity. Depending on which capacity is zero, two scenarios can occur (see also Fig. \ref{fig: ZF-RE model}):
\begin{enumerate}
    \item $C_{21}=0$ termed as zero-forcing scenario.
    \item $C_{12}=0$ termed as relaying scenario.
\end{enumerate}
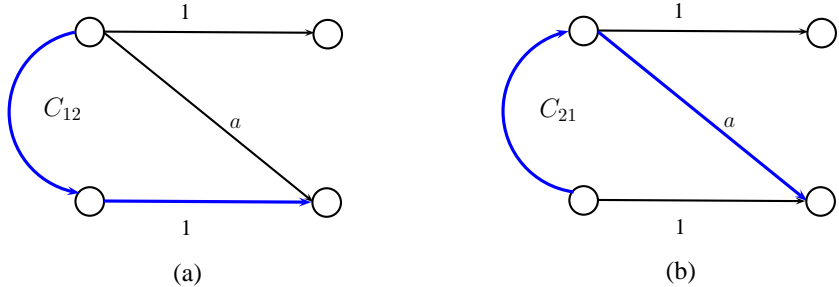
\begin{figure}
\begin{center}
\scalebox{.7}
{
\begin{pspicture}(0,-2.7535937)(15.7225,2.7535937)
\pscircle[linewidth=0.04,dimen=outer](10.91125,2.162344){0.2912501}
\psline[linewidth=0.04cm,arrowsize=0.05291667cm 2.0,arrowlength=1.4,arrowinset=0.4]{->}(11.2025,2.171094)(15.18,2.1492188)
\psline[linewidth=0.06cm,linecolor=blue,arrowsize=0.05291667cm 2.0,arrowlength=1.4,arrowinset=0.4]{->}(11.1825,2.151094)(15.16,-1.0707812)
\usefont{T1}{ptm}{m}{n}
\rput(12.726093,2.5510938){\large 1}
\usefont{T1}{ptm}{m}{n}
\rput(12.746094,-1.5489062){\large 1}
\usefont{T1}{ptm}{m}{n}
\rput(13.680469,0.4310939){\large $a$}
\pscircle[linewidth=0.04,dimen=outer](15.41125,-1.077656){0.2912501}
\pscircle[linewidth=0.04,dimen=outer](15.431251,2.122344){0.2912501}
\pscircle[linewidth=0.04,dimen=outer](10.91125,-1.057656){0.2912501}
\psline[linewidth=0.04cm,arrowsize=0.05291667cm 2.0,arrowlength=1.4,arrowinset=0.4]{->}(11.1825,-1.0489061)(15.12,-1.0707812)
\pscircle[linewidth=0.04,dimen=outer](1.5312501,2.142344){0.2912501}
\psline[linewidth=0.04cm,arrowsize=0.05291667cm 2.0,arrowlength=1.4,arrowinset=0.4]{->}(1.8225,2.151094)(5.8,2.1292188)
\psline[linewidth=0.04cm,arrowsize=0.05291667cm 2.0,arrowlength=1.4,arrowinset=0.4]{->}(1.8025001,2.1310937)(5.78,-1.0907812)
\usefont{T1}{ptm}{m}{n}
\rput(3.3460937,2.5310938){\large 1}
\usefont{T1}{ptm}{m}{n}
\rput(3.3660936,-1.5689062){\large 1}
\usefont{T1}{ptm}{m}{n}
\rput(4.300469,0.4110939){\large $a$}
\pscircle[linewidth=0.04,dimen=outer](6.03125,-1.097656){0.2912501}
\pscircle[linewidth=0.04,dimen=outer](6.05125,2.102344){0.2912501}
\pscircle[linewidth=0.04,dimen=outer](1.5312501,-1.077656){0.2912501}
\psline[linewidth=0.06cm,linecolor=blue,arrowsize=0.05291667cm 2.0,arrowlength=1.4,arrowinset=0.4]{->}(1.8025,-1.0689061)(5.74,-1.0907812)
\rput{-180.0}(21.86,1.2784375){\psarc[linewidth=0.06,linecolor=blue,arrowsize=0.05291667cm 2.0,arrowlength=1.4,arrowinset=0.4]{<-}(10.93,0.63921875){1.55}{281.30994}{82.278175}}
\rput{-180.0}(3.1,1.2384375){\psarc[linewidth=0.06,linecolor=blue,arrowsize=0.05291667cm 2.0,arrowlength=1.4,arrowinset=0.4]{->}(1.55,0.61921877){1.55}{281.30994}{82.278175}}
\usefont{T1}{ptm}{m}{n}
\rput(1.0153126,0.68921876){\Large $C_{12}$}
\usefont{T1}{ptm}{m}{n}
\rput(10.435312,0.6692188){\Large $C_{21}$}
\usefont{T1}{ptm}{m}{n}
\rput(3.3903124,-2.4707813){\Large (a)}
\usefont{T1}{ptm}{m}{n}
\rput(12.760312,-2.4307814){\Large (b)}
\end{pspicture}
}
\caption{Unidirectional Cooperation: (a) Zero-Forcing Scenario, (b) Relaying Scenario. The blue color represents the route of zero-forcing/ relaying.}\label{fig: ZF-RE model}
\vspace{-13pt}
\end{center}
\end{figure}
\vspace{-2pt}
\subsection{Zero-Forcing Scenario ($C_{21}=0$)}\label{subsec: ZF}
In this scenario, Transmitter 2 utilizes $C_{12}$ to cancel the known part of the interference (Fig. \ref{fig: ZF-RE model} (a)). The rest of the signaling is similar to the one proposed for the conventional GIC \cite{EtkinIT07}. In other words, Encoder 1 makes use of three independent codebooks\footnote{Through out the paper, it is assumed that all of the employed codebooks are Gaussian and independent of each other.}, namely private codebooks $C_{1}^{p}$ and $C_{1}^{z}$, and common codebook $C_{1}^{c}$, with corresponding codewords $X_{1p}, X_{1z}$, and $X_{1c}$. Encoder 2 uses two private codebooks $C_{2}^{p}$ and $C_{1}^{z}$. $X_{1z}$ is available at both transmitters via the cooperative link and zero-forced at the second receiver. Since Transmitter 2 does not cause any interference on Receiver 1, there is no need to include a common codebook for User 2. The transmit signals are represented as follows:
\begin{align}\label{eq: sign_UC_ZF}
\begin{split}
X_1&=X_{1p}+X_{1c}+X_{1z},\\
X_2&= X_{2p} -a X_{1z}.
\end{split}
\end{align}
Decoders decide on the codeword indices $i, j, k,$ and $l$ according to:
\begin{itemize}
  \item Decoder 1: $\lp X_{1p}(i), X_{1z}(j),X_{1c}(k),Y_1\rp\in A_{\epsilon}^{(n)}$,
  \item Decoder 2: $\lp X_{1c}(k),X_{2p}(l),Y_2\rp\in A_{\epsilon}^{(n)}$.
\end{itemize}
Appendix \ref{app: Gap_UC_ZF} shows the rate region described by (\ref{eq: Ach_ZIC_UC_ZF_R1})-(\ref{eq: Ach_ZIC_UC_ZF_R1pR2}) is achievable:
\begin{align}
R_1&\!\le\!\min\{\C(P_1),\C(P_{1p}\!+\!P_{1c})\!+\!C_{12}\}\label{eq: Ach_ZIC_UC_ZF_R1}\\
R_{2}&\!\le\!\C(\frac{P_{2p}}{d})\label{eq: Ach_ZIC_UC_ZF_R2}\\
R_{1}+R_{2}&\!\le\!\min\{\C(P_{1p}\!+\!P_{1z}),\C(P_{1p})\!+\!C_{12}\}\!+\!\C(\frac{a^2P_{1c}\!+\!P_{2p}}{d}),\label{eq: Ach_ZIC_UC_ZF_R1pR2}
\end{align}
where
\begin{equation}\label{eq: d}
   d\triangleq 1+a^2P_{1p}.
\end{equation}
The following observations are utilized to attain a suitable power allocation for different codewords:
\begin{itemize}
  \item Since $X_{1p}$ is treated as noise at Receiver 2, we set $P_{1p}\!=\!\Ppr$ in order to receive $X_{1p}$ at the level of the Gaussian noise at Receiver 2 \cite{EtkinIT07}.
  \item To make $R_2$ close to $R_2^{\text{up}}$, \emph{i.e.}, $\C(P_2)$, we impose the constraint $\frac{P_2}{2}\!\le\!P_{2p}$.
  \item To make $R_1$ in (\ref{eq: Ach_ZIC_UC_ZF_R1}) close to $R_1^{\text{up}}$, \emph{i.e.}, $\C(P_1)$, we enforce $\frac{P_1-P_{1p}}{2}\!\le\!P_{1c}$. This requirement is more pronounced when $C_{12}=0$.
\end{itemize}
For the last two items, the factor $2$ in the denominators ensures a maximum loss of 0.5 bit compared to the case of $P_{2p}=P_2$, and $P_{1c}=P_1-P_{1p}$. Therefore, to satisfy the above constraints, we select $P_{1z}$ according to:
\vspace{1pt}
\begin{align}\label{eq: P1z}
P_{1z}=\min\lp \frac{P_1-P_{1p}}{2}, \frac{P_2}{2a^2} \rp. 
\end{align}
\begin{lemma}\label{lem: Gap_UC_ZF}
The preceding achievable scheme is ${\cal R}(0.5, 1)$ achievable.
\end{lemma}
\begin{IEEEproof}
See Appendix \ref{app: Gap_UC_ZF}.
\end{IEEEproof}
\subsection{Relaying Scenario ($C_{12}=0$)}\label{subsec: Re}
Here, it is assumed that $C_{12}\!=\!0$. In this scenario, $C_{21}$ is employed to help Encoder 1 relay some information for User 2 (Fig. \ref{fig: ZF-RE model} (b)). Based on the relaying capability (the relay power $a^2P_1$ and the relay to destination channel gain $a$), three cases are recognized in this setup and for each case, a different achievable scheme is proposed:
\begin{itemize}
  \item Non Cooperative Case: $a^2P_1\leq P_2+1$,
  \item Common Cooperative Case: $P_2+1< a^2P_1$, $a^2\leq P_2+1$,
  \item Private-Common Cooperative Case: $P_2+1< a^2P_1$, $P_2+1 < a^2$.
\end{itemize}
Before we continue to describe the achievable schemes, we stress that, throughout the paper, we aim to keep the achievable schemes simple, at the expense of a slight increase in the gap from the upper bound.
\subsubsection{Non Cooperative Case: $a^2P_1\leq P_2+1$}\label{subsubsec: nc_Re}
In this case, the cooperative link is not used because it can at most enhance $R_2$ by one bit. The signaling is similar to the HK signaling developed for the GIC:
\begin{align}
\begin{split}
X_1&= X_{1p}+X_{1c},\\
X_2&=X_{2p}.
\end{split}
\end{align}
We set $P_{1c}=P_1-P_{1p}$ with $P_{1p}=\frac{1}{a^2}$ for $a\le 1$, and $P_{1p}=0$ for $1<a$.
The decoding rules are:
\begin{itemize}
  \item Decoder 1: $\lp X_{1p}(i), X_{1c}(j), Y_1\rp\in A_{\epsilon}^{(n)}$,
  \item Decoder 2: $\lp X_{1c}(j), X_{2p}(k), Y_2\rp\in A_{\epsilon}^{(n)}$,
\end{itemize}
which lead to the ${\cal{R}}(0,1.5)$ achievable region below:
\vspace{-4pt}
\begin{align}
R_1&\leq \C(P_1)\label{eq: R1_Re_nc}\\
R_2&\leq \C(\frac{P_2}{d})\label{eq: R2_Re_nc}\\
R_1+R_2&\leq \C(P_{1p})+\C(\frac{a^2(P_1-P_{1p})+P_2}{d}).\label{eq: R1pR2_Re_nc}
\end{align}
See Appendix \ref{app: Ach_Gap_UC_Re_1} for details.
\subsubsection{Common Cooperative Case: $P_2+1 < a^2P_1, a^2 \leq P_2+1$}
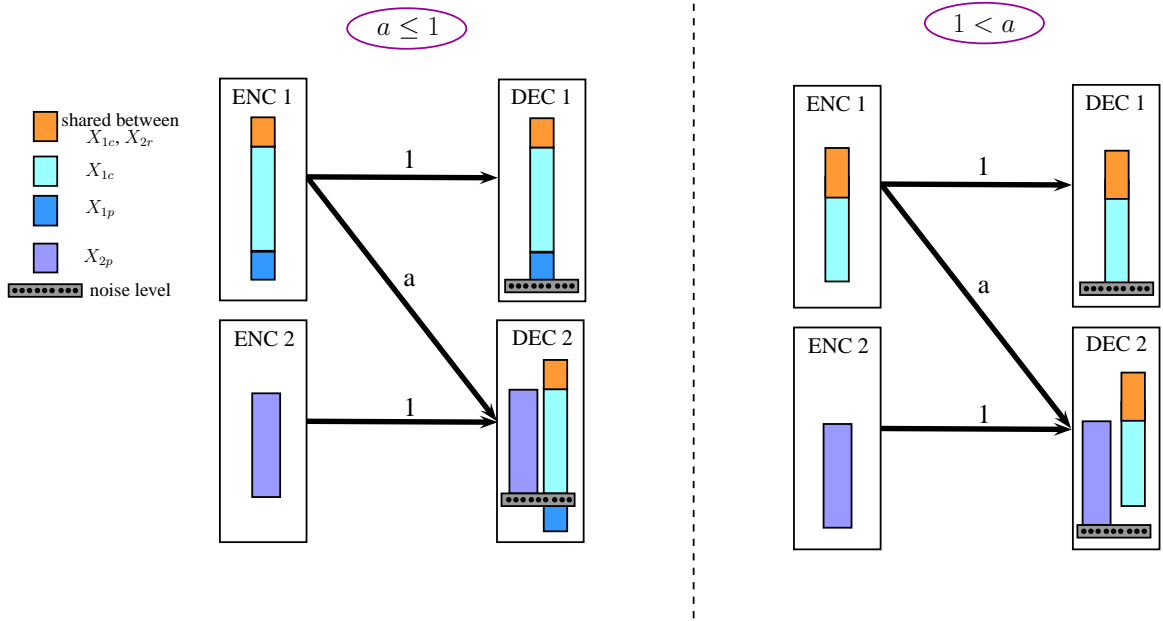
\begin{figure}
\begin{center}
\scalebox{.6}
{
\begin{pspicture}(0,-6.854375)(25.54,6.834375)
\definecolor{color1b}{rgb}{0.6,0.6,1.0}
\definecolor{color2b}{rgb}{0.2,0.6,1.0}
\definecolor{color3b}{rgb}{0.6,1.0,1.0}
\definecolor{color886b}{rgb}{1.0,0.6,0.2}
\definecolor{color21b}{rgb}{0.6,0.6,0.6}
\definecolor{color80}{rgb}{0.6,0.0,0.6}
\psline[linewidth=0.04cm](11.095781,-4.0453124)(12.275781,-4.0853124)
\psframe[linewidth=0.04,dimen=outer,fillstyle=solid,fillcolor=color1b](11.755781,-1.7453125)(11.095781,-4.0853124)
\psframe[linewidth=0.04,dimen=outer,fillstyle=solid,fillcolor=color2b](12.115781,1.2946877)(11.555781,0.6346876)
\psframe[linewidth=0.04,dimen=outer,fillstyle=solid,fillcolor=color3b](12.115781,3.6256251)(11.555781,1.2746876)
\psframe[linewidth=0.04,dimen=outer,fillstyle=solid,fillcolor=color886b](12.115781,4.2746882)(11.555781,3.5856252)
\psframe[linewidth=0.04,dimen=outer,fillstyle=solid,fillcolor=color2b](12.415781,-4.0653124)(11.855781,-4.925625)
\psframe[linewidth=0.04,dimen=outer,fillstyle=solid,fillcolor=color3b](12.415781,-1.7343748)(11.855781,-4.0853124)
\psframe[linewidth=0.04,dimen=outer,fillstyle=solid,fillcolor=color886b](12.415781,-1.0853122)(11.855781,-1.7743748)
\psline[linewidth=0.04cm,linestyle=dashed,dash=0.16cm 0.16cm](15.2,6.794375)(15.196875,-6.834375)
\psframe[linewidth=0.04,dimen=outer,fillstyle=solid,fillcolor=color1b](24.455782,-2.4453125)(23.795782,-4.7853127)
\psframe[linewidth=0.04,dimen=outer,fillstyle=solid,fillcolor=color3b](25.21578,-2.0143747)(24.655783,-4.3653126)
\psframe[linewidth=0.04,dimen=outer,fillstyle=solid,fillcolor=color886b](25.21578,-1.3653122)(24.655783,-2.4743745)
\usefont{T1}{ptm}{m}{n}
\rput(21.638748,6.349688){\LARGE $1< a$}
\usefont{T1}{ptm}{m}{n}
\rput(8.87875,6.249688){\LARGE $a\leq 1$}
\psframe[linewidth=0.04,dimen=outer,fillstyle=solid,fillcolor=color3b](24.855782,2.905625)(24.295782,0.5546876)
\psframe[linewidth=0.04,dimen=outer,fillstyle=solid,fillcolor=color886b](24.855782,3.5546875)(24.295782,2.4456253)
\psframe[linewidth=0.04,dimen=outer,fillstyle=solid,fillcolor=color886b](1.1157811,4.414688)(0.5557812,3.725625)
\psframe[linewidth=0.04,dimen=outer,fillstyle=solid,fillcolor=color3b](1.1157811,3.425625)(0.5557812,2.7343748)
\psframe[linewidth=0.04,dimen=outer,fillstyle=solid,fillcolor=color2b](1.1157811,2.554375)(0.5557812,1.8546876)
\usefont{T1}{ptm}{m}{n}
\rput(2.4915624,3.820625){\large $X_{1c}$, $X_{2r}$}
\usefont{T1}{ptm}{m}{n}
\rput(2.5,4.220625){\large shared between}
\psframe[linewidth=0.04,dimen=outer,fillstyle=solid,fillcolor=color21b](1.66,0.5943749)(0.0,0.2743749)
\psdots[dotsize=0.12](1.18,0.4343749)
\psdots[dotsize=0.12](1.34,0.4343749)
\psdots[dotsize=0.12](1.5,0.4343749)
\psdots[dotsize=0.12](0.98,0.4343749)
\psdots[dotsize=0.12](0.82,0.4343749)
\psdots[dotsize=0.12](0.66,0.4343749)
\usefont{T1}{ptm}{m}{n}
\rput(2.720625,0.4493749){\large noise level}
\usefont{T1}{ptm}{m}{n}
\rput(2.0415626,3.080625){\large $X_{1c}$}
\psframe[linewidth=0.04,dimen=outer,fillstyle=solid,fillcolor=color1b](1.12,1.4946874)(0.55578125,0.8143749)
\usefont{T1}{ptm}{m}{n}
\rput(2.0515625,2.2406251){\large $X_{1p}$}
\usefont{T1}{ptm}{m}{n}
\rput(2.0115623,1.1406252){\large $X_{2p}$}
\psdots[dotsize=0.12](0.16,0.4343749)
\psdots[dotsize=0.12](0.32,0.4343749)
\psdots[dotsize=0.12](0.48,0.4343749)
\psframe[linewidth=0.04,dimen=outer,fillstyle=solid,fillcolor=color21b](12.66,0.6943749)(11.0,0.3743749)
\psdots[dotsize=0.12](12.18,0.5343749)
\psdots[dotsize=0.12](12.34,0.5343749)
\psdots[dotsize=0.12](12.5,0.5343749)
\psdots[dotsize=0.12](11.98,0.5343749)
\psdots[dotsize=0.12](11.82,0.5343749)
\psdots[dotsize=0.12](11.66,0.5343749)
\psdots[dotsize=0.12](11.16,0.5343749)
\psdots[dotsize=0.12](11.32,0.5343749)
\psdots[dotsize=0.12](11.48,0.5343749)
\psframe[linewidth=0.04,dimen=outer,fillstyle=solid,fillcolor=color21b](25.4,0.6343749)(23.74,0.3143749)
\psdots[dotsize=0.12](24.92,0.4743749)
\psdots[dotsize=0.12](25.08,0.4743749)
\psdots[dotsize=0.12](25.24,0.4743749)
\psdots[dotsize=0.12](24.72,0.4743749)
\psdots[dotsize=0.12](24.56,0.4743749)
\psdots[dotsize=0.12](24.4,0.4743749)
\psdots[dotsize=0.12](23.9,0.4743749)
\psdots[dotsize=0.12](24.06,0.4743749)
\psdots[dotsize=0.12](24.22,0.4743749)
\psframe[linewidth=0.04,dimen=outer,fillstyle=solid,fillcolor=color21b](25.34,-4.745625)(23.68,-5.065625)
\psdots[dotsize=0.12](24.86,-4.905625)
\psdots[dotsize=0.12](25.02,-4.905625)
\psdots[dotsize=0.12](25.18,-4.905625)
\psdots[dotsize=0.12](24.66,-4.905625)
\psdots[dotsize=0.12](24.5,-4.905625)
\psdots[dotsize=0.12](24.34,-4.905625)
\psdots[dotsize=0.12](23.84,-4.905625)
\psdots[dotsize=0.12](24.0,-4.905625)
\psdots[dotsize=0.12](24.16,-4.905625)
\psframe[linewidth=0.04,dimen=outer,fillstyle=solid,fillcolor=color21b](12.58,-4.045625)(10.92,-4.365625)
\psdots[dotsize=0.12](12.12,-4.205625)
\psdots[dotsize=0.12](12.28,-4.205625)
\psdots[dotsize=0.12](12.44,-4.205625)
\psdots[dotsize=0.12](11.92,-4.205625)
\psdots[dotsize=0.12](11.76,-4.205625)
\psdots[dotsize=0.12](11.6,-4.205625)
\psdots[dotsize=0.12](11.1,-4.205625)
\psdots[dotsize=0.12](11.26,-4.205625)
\psdots[dotsize=0.12](11.42,-4.205625)
\psframe[linewidth=0.04,dimen=outer](12.82,5.134375)(10.86,0.1743749)
\usefont{T1}{ptm}{m}{n}
\rput(11.833438,4.734375){\Large DEC 1}
\psframe[linewidth=0.04,dimen=outer](12.78,-0.2056251)(10.82,-5.165625)
\usefont{T1}{ptm}{m}{n}
\rput(11.851874,-0.6056251){\Large DEC 2}
\psellipse[linewidth=0.04,linecolor=color80,dimen=outer](8.84,6.244375)(1.34,0.47)
\psellipse[linewidth=0.04,linecolor=color80,dimen=outer](21.64,6.364375)(1.34,0.47)
\psframe[linewidth=0.04,dimen=outer](25.54,5.014375)(23.58,0.0543749)
\usefont{T1}{ptm}{m}{n}
\rput(24.553438,4.594375){\Large DEC 1}
\psframe[linewidth=0.04,dimen=outer](25.54,-0.3656251)(23.58,-5.3256254)
\usefont{T1}{ptm}{m}{n}
\rput(24.571875,-0.7456251){\Large DEC 2}
\psframe[linewidth=0.04,dimen=outer,fillstyle=solid,fillcolor=color1b](6.0557814,-1.8253125)(5.3957815,-4.1653123)
\psframe[linewidth=0.04,dimen=outer,fillstyle=solid,fillcolor=color2b](5.935781,1.3146877)(5.375781,0.6546876)
\psframe[linewidth=0.04,dimen=outer,fillstyle=solid,fillcolor=color3b](5.935781,3.645625)(5.375781,1.2946877)
\psframe[linewidth=0.04,dimen=outer,fillstyle=solid,fillcolor=color886b](5.935781,4.294688)(5.375781,3.6056252)
\psframe[linewidth=0.04,dimen=outer](6.64,5.154375)(4.68,0.1943749)
\usefont{T1}{ptm}{m}{n}
\rput(5.6525,4.734375){\Large ENC 1}
\psframe[linewidth=0.04,dimen=outer](6.64,-0.2056251)(4.68,-5.165625)
\usefont{T1}{ptm}{m}{n}
\rput(5.6909375,-0.6256251){\Large ENC 2}
\psline[linewidth=0.12cm,arrowsize=0.05291667cm 2.0,arrowlength=1.4,arrowinset=0.4]{->}(6.62,2.954375)(10.86,2.934375)
\psline[linewidth=0.12cm,arrowsize=0.05291667cm 2.0,arrowlength=1.4,arrowinset=0.4]{->}(6.62,-2.465625)(10.86,-2.485625)
\psline[linewidth=0.12cm,arrowsize=0.05291667cm 2.0,arrowlength=1.4,arrowinset=0.4]{->}(6.66,2.914375)(10.84,-2.465625)
\usefont{T1}{ptm}{m}{n}
\rput(8.898281,3.354375){\LARGE 1}
\usefont{T1}{ptm}{m}{n}
\rput(8.901875,0.674375){\LARGE a}
\usefont{T1}{ptm}{m}{n}
\rput(8.918282,-2.145625){\LARGE 1}
\psframe[linewidth=0.04,dimen=outer,fillstyle=solid,fillcolor=color1b](18.715782,-2.5053124)(18.055782,-4.8453126)
\psframe[linewidth=0.04,dimen=outer,fillstyle=solid,fillcolor=color3b](18.65578,2.9656253)(18.095781,0.6146877)
\psframe[linewidth=0.04,dimen=outer,fillstyle=solid,fillcolor=color886b](18.65578,3.6146884)(18.095781,2.474375)
\psframe[linewidth=0.04,dimen=outer](19.36,4.994375)(17.4,0.0343749)
\usefont{T1}{ptm}{m}{n}
\rput(18.3725,4.574375){\Large ENC 1}
\psframe[linewidth=0.04,dimen=outer](19.36,-0.3656251)(17.4,-5.3256254)
\usefont{T1}{ptm}{m}{n}
\rput(18.410936,-0.7856251){\Large ENC 2}
\psline[linewidth=0.12cm,arrowsize=0.05291667cm 2.0,arrowlength=1.4,arrowinset=0.4]{->}(19.34,2.794375)(23.58,2.774375)
\psline[linewidth=0.12cm,arrowsize=0.05291667cm 2.0,arrowlength=1.4,arrowinset=0.4]{->}(19.34,-2.625625)(23.58,-2.645625)
\psline[linewidth=0.12cm,arrowsize=0.05291667cm 2.0,arrowlength=1.4,arrowinset=0.4]{->}(19.38,2.754375)(23.56,-2.625625)
\usefont{T1}{ptm}{m}{n}
\rput(21.61828,3.194375){\LARGE 1}
\usefont{T1}{ptm}{m}{n}
\rput(21.621876,0.514375){\LARGE a}
\usefont{T1}{ptm}{m}{n}
\rput(21.63828,-2.305625){\LARGE 1}
\end{pspicture}
}
\caption{LDM for GZIC with $C_{12}=0$: Common Cooperative Case. The received power level at each receiver is shown for $a\le 1$, and $1 < a$ situations. Different colors represent different signals. The signals with the same power level should be decoded jointly.}\label{fig: det_zu_re_cc}
\end{center}
\end{figure}
\vspace{-3pt}
For this case, in addition to the signals transmitted in the non cooperative case, User 1 relays some data communicated over the cooperative link for User 2. The signaling is as follows:
\vspace{-4pt}
\begin{align}
\begin{split}
X_1&= X_{1p}+X_{1c}+X_{2r},\\
X_2&=X_{2p},
\end{split}
\end{align}
where all signals are independent of each other.
To find out how to treat $X_{2r}$ at Receiver 1 and how to allocate the Transmitter 1's power between $X_{1c}$ and $X_{2r}$, the deterministic model shown in Fig. \ref{fig: det_zu_re_cc}, is used. The model demonstrates the power level interaction of the interfering signals according to the channel parameters. Since, $X_{1p}$ is received below the noise level of Decoder 2, it is considered as noise at that receiver. When two signals are at the same power level at a decoder, they need to be decoded jointly at that receiver. It is clear that there is not much benefit in having two signals with the same power level intended for the same receiver.\footnote{One can infer that in the non cooperative case, the relayed signal is not required (in the constant gap sense). This is because in the corresponding LDM (not shown in the paper), $X_{2p}$ and $X_{2r}$ would share the same power level.} Having said that, it is noticed that the relayed signal has to share its power level with the common signal of User 1 to not limit the User 1's rate by $P_2$ (see power levels at Dec 2 in Fig. \ref{fig: det_zu_re_cc}). Therefore, the relayed signal is considered as a common signal, and consequently decoded at both receivers.
The decoding rules are:
\begin{itemize}
  \item Decoder 1: $\lp X_{1p}(i), X_{1c}(j), X_{2r}(k), Y_1\rp\in A_{\epsilon}^{(n)}$,
  \item Decoder 2: $\lp X_{1c}(j), X_{2r}(k), X_{2p}(l), Y_2\rp\in A_{\epsilon}^{(n)}$.
\end{itemize}
For allocating power among the codebooks, the LDM suggests $X_{2r}$ and $X_{1c}$ to have the same power level. Hence, we set
$P_{2p}\!=\!P_2$, $P_{2r}\!=\!P_{1c}\!=\!\frac{P_1-P_{1p}}{2}$, where $P_{1p}\!=\!\frac{1}{a^2}$ for $a\!\le\!1$,
and $P_{1p}\!=\!0$ otherwise.
\begin{lemma}\label{lem: Ach_Gap_UC_Re_2}
The following region is ${\cal R}(0.5,1.5)$, and ${\cal R}(0.5,0.5)$ achievable for $a\!\le\!1$, and $1\!<\!a$ cases, respectively:\footnote{For $1<a$ the second condition for $R_2$ is redundant.}
\vspace{-3pt}
\begin{align}
R_{1}&\le \C(\frac{P_1+P_{1p}}{2})\label{eq: Rm1_1_c}\\
R_{2}&\le \C(\frac{P_{2p}}{d})+C_{21}\label{eq: Rm2_1_c}\\
R_{2}&\le \C(\frac{a^2(P_{1}-P_{1p})+2P_{2p}}{2d})\label{eq: Rm2_2_c}\\
R_{1}+R_{2}&\le \C(P_{1p})+\C(\frac{a^2 (P_1-P_{1p})+P_{2p}}{d})-\frac{1}{2}\label{eq: Rm1pR2_3_c},
\end{align}
where $d$ is defined in (\ref{eq: d}).
\end{lemma}
\begin{IEEEproof}
See Appendix \ref{app: Ach_Gap_UC_Re_2}.
\end{IEEEproof}
\subsubsection{Private-Common Cooperative Case: $P_2\!+\!1\!<\!a^2P_1, P_2\!+\!1\!<\!a^2$}\label{sec: PCCC}
\begin{figure}
\begin{center}
\scalebox{.6}
{
\begin{pspicture}(0,-5.29)(12.5,5.29)
\definecolor{color1988b}{rgb}{0.6,1.0,1.0}
\definecolor{color1989b}{rgb}{1.0,0.6,0.2}
\definecolor{color1990b}{rgb}{0.6,0.6,1.0}
\definecolor{color1996b}{rgb}{0.6,0.6,0.6}
\psframe[linewidth=0.04,dimen=outer,fillstyle=solid,fillcolor=color1988b](11.890938,3.88625)(11.330937,2.3353126)
\psframe[linewidth=0.04,dimen=outer,fillstyle=solid,fillcolor=color1989b](11.890938,2.0753126)(11.330937,1.3862498)
\psframe[linewidth=0.04,dimen=outer,fillstyle=solid,fillcolor=color1990b](11.910938,-2.54375)(11.250937,-4.8846874)
\psframe[linewidth=0.04,dimen=outer,fillstyle=solid,fillcolor=color1988b](12.390938,-0.13375)(11.830938,-1.6846876)
\psframe[linewidth=0.04,dimen=outer,fillstyle=solid,fillcolor=color1989b](1.1157811,3.290313)(0.5557812,2.60125)
\psframe[linewidth=0.04,dimen=outer,fillstyle=solid,fillcolor=color1988b](1.135781,4.32125)(0.5757812,3.63)
\usefont{T1}{ptm}{m}{n}
\rput(2.806875,3.6412504){\Large $X_{1c}$, $X_{2r_c}$}
\usefont{T1}{ptm}{m}{n}
\rput(2.6771874,4.14125){\Large shared between}
\psframe[linewidth=0.04,dimen=outer,fillstyle=solid,fillcolor=color1996b](1.66,1.37)(0.0,1.05)
\psdots[dotsize=0.12](1.18,1.21)
\psdots[dotsize=0.12](1.34,1.21)
\psdots[dotsize=0.12](1.5,1.21)
\psdots[dotsize=0.12](0.98,1.21)
\psdots[dotsize=0.12](0.82,1.21)
\psdots[dotsize=0.12](0.66,1.21)
\usefont{T1}{ptm}{m}{n}
\rput(2.7909374,1.23){\Large noise level}
\usefont{T1}{ptm}{m}{n}
\rput(2.2768753,2.88125){\Large $X_{2r_p}$}
\psframe[linewidth=0.04,dimen=outer,fillstyle=solid,fillcolor=color1990b](1.12,2.2703123)(0.55578125,1.59)
\usefont{T1}{ptm}{m}{n}
\rput(2.136875,1.92125){\Large $X_{2p}$}
\psdots[dotsize=0.12](0.16,1.21)
\psdots[dotsize=0.12](0.32,1.21)
\psdots[dotsize=0.12](0.48,1.21)
\psframe[linewidth=0.04,dimen=outer,fillstyle=solid,fillcolor=color1996b](12.3,-4.79)(10.64,-5.11)
\psdots[dotsize=0.12](11.84,-4.95)
\psdots[dotsize=0.12](12.0,-4.95)
\psdots[dotsize=0.12](12.16,-4.95)
\psdots[dotsize=0.12](11.64,-4.95)
\psdots[dotsize=0.12](11.48,-4.95)
\psdots[dotsize=0.12](11.32,-4.95)
\psdots[dotsize=0.12](10.82,-4.95)
\psdots[dotsize=0.12](10.98,-4.95)
\psdots[dotsize=0.12](11.14,-4.95)
\psframe[linewidth=0.04,dimen=outer](12.5,5.29)(10.54,1.21)
\usefont{T1}{ptm}{m}{n}
\rput(11.520157,4.83){\Large DEC 1}
\psframe[linewidth=0.04,dimen=outer](12.5,0.65)(10.54,-5.27)
\usefont{T1}{ptm}{m}{n}
\rput(11.517812,0.21){\Large DEC 2}
\psframe[linewidth=0.04,dimen=outer,fillstyle=solid,fillcolor=color1996b](12.4,2.37)(10.74,2.05)
\psdots[dotsize=0.12](11.92,2.21)
\psdots[dotsize=0.12](12.08,2.21)
\psdots[dotsize=0.12](12.24,2.21)
\psdots[dotsize=0.12](11.72,2.21)
\psdots[dotsize=0.12](11.56,2.21)
\psdots[dotsize=0.12](11.4,2.21)
\psdots[dotsize=0.12](10.9,2.21)
\psdots[dotsize=0.12](11.06,2.21)
\psdots[dotsize=0.12](11.22,2.21)
\psframe[linewidth=0.04,dimen=outer,fillstyle=solid,fillcolor=color1989b](12.390938,-1.6446874)(11.830938,-2.33375)
\psframe[linewidth=0.04,dimen=outer,fillstyle=solid,fillcolor=color1988b](6.2709374,3.8662498)(5.7109375,2.09)
\psframe[linewidth=0.04,dimen=outer,fillstyle=solid,fillcolor=color1989b](6.2709374,2.13)(5.7109375,1.3662498)
\psframe[linewidth=0.04,dimen=outer,fillstyle=solid,fillcolor=color1990b](6.290938,-2.56375)(5.630937,-4.9046874)
\psframe[linewidth=0.04,dimen=outer](6.88,5.27)(4.92,1.19)
\usefont{T1}{ptm}{m}{n}
\rput(5.899219,4.81){\Large ENC 1}
\psframe[linewidth=0.04,dimen=outer](6.9,-1.11)(4.94,-5.29)
\usefont{T1}{ptm}{m}{n}
\rput(5.896874,-1.83){\Large ENC 2}
\psline[linewidth=0.12cm,arrowsize=0.05291667cm 2.0,arrowlength=1.4,arrowinset=0.4]{->}(6.86,3.23)(10.58,3.23)
\psline[linewidth=0.12cm,arrowsize=0.05291667cm 2.0,arrowlength=1.4,arrowinset=0.4]{->}(6.86,-2.39)(10.58,-2.39)
\psline[linewidth=0.12cm,arrowsize=0.05291667cm 2.0,arrowlength=1.4,arrowinset=0.4]{->}(6.92,3.19)(10.58,-2.43)
\usefont{T1}{ptm}{m}{n}
\rput(8.61125,3.755){\huge 1}
\usefont{T1}{ptm}{m}{n}
\rput(8.71125,-2.865){\huge 1}
\usefont{T1}{ptm}{m}{n}
\rput(8.922188,0.755){\huge a}
\end{pspicture}
}
\caption{LDM for GZIC with $C_{12}=0$: Private-Common Cooperative Case.}\label{fig: det_zu_re_pcc}
\end{center}
\end{figure}
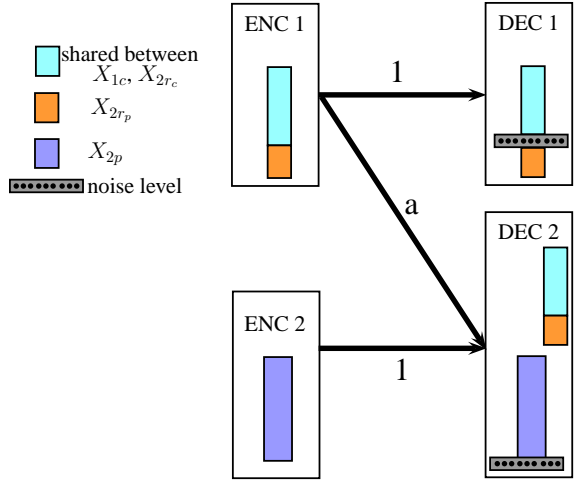
For this case, the corresponding LDM shown in Fig. \ref{fig: det_zu_re_pcc} illustrates that since $a$ is quite large, Transmitter 1 can spend a small amount of power (\emph{i.e.}, less than 1 unit) to relay the signal $X_{2r_p}$ for User 2 without a decoding requirement at Receiver 1. This technique can be considered as the counterpart of the transmission at the noise level originally proposed for the conventional GIC, and used in this paper for $a\le 1$ configurations. This model also suggests that a common signal, named $X_{2r_c}$, needs to be relayed for User 2, similar to the $a^2 \le P_2+1$ case. Therefore, the following signaling is used:
\begin{align}
\begin{split}
X_1&=X_{1c}+X_{2r_p}+X_{2r_c},\\
X_2&=X_{2p},
\end{split}
\end{align}
where $X_{1c}, X_{2r_c}$ are common signals and $X_{2p}, X_{2r_p}$ are private signals for User 2.
The decoding rules are:
\begin{itemize}
  \item Decoder 1: $\lp X_{1c}(i), X_{2r}(j), Y_1\rp\in A_{\epsilon}^{(n)}$,
  \item Decoder 2: $\lp X_{1c}(i), X_{2r}(j), X_{2r_p}(k), X_{2p}(l), Y_2\rp\in A_{\epsilon}^{(n)}$.
\end{itemize}
To avoid the complexity of jointly decoding four signals at Receiver 2, first $X_{1c}, X_{2r_c}, X_{2r_p}$ are jointly decoded assuming $X_{2p}$ as noise and then $X_{2p}$ is decoded.
To appropriately allocate the power of Transmitter 1 to different codebooks, the LDM suggests to provide $\min\{1, P_1\}$ amount of power for $X_{2r_p}$ transmission. Since this signal is received below the noise level of Receiver 1, there is no much cost associated with its transmission, and therefore, it can be considered as the first signal to get its share of power. The rest of the power is equally distributed between $X_{1c}$ and $X_{2r_c}$ as the case for $a^2 \le P_2+1$. One question to answer is how to divide the capacity $C_{21}$ between $X_{2r_p}$ and $X_{2r_c}$? Again, we give the priority to $X_{2r_p}$, and name its share of the cooperative capacity, $C'$. Below, $P_1\!\leq\!1$, and $1\!<\!P_1$ cases are considered, respectively.

If $P_1\!\leq\!1$, we set $P_{2r_p}\!=\!P_1$, $P_{1c}\!=\!P_{2r_c}\!=\!0$, and $C'=C_{21}$. It is easy to show $(
0, \min\{\C(P_2)\!+\!C_{21}, \C(a^2 P_1\!+\!P_2)\})$ is achievable. Comparing this rate pair with the upper bounds (\ref{eq: Ru1_1})-(\ref{eq: Ru2_2}) proves that ${\cal R}(0.5,0.5)$ is achievable.

Now, it is assumed that $1\!<\!P_1$. We set $P_{2r_p}\!=\!1$, $P_{1c}\!=\!$ $P_{2r_c}\!=\!\frac{P_1\!-\!1}{2}$, and
$P_{2p}\!=\!P_2$. The decoding error analysis of Appendix \ref{app: Ach_Gap_UC_Re_3} states that $R_{2r_p}\le \C(\frac{a^2}{P_2+1})$, and therefore, we set $C'=\min\{\C(\frac{a^2}{P_2+1}),C_{21}\}$.

\begin{lemma}\label{lem: Ach_Gap_UC_Re_3}
The following region is ${\cal R}(1,0.5)$ achievable:
\begin{align}\label{eq: Ach_Re_3}
\begin{split}
R_1&\le \C(\frac{P_1-1}{4})\\
R_2&\le \C(P_2)+C_{21}\\
R_1+R_2&\le \C(a^2 P_1+P_2)-\frac{1}{2}.
\end{split}
\end{align}
\end{lemma}
\begin{IEEEproof}
See Appendix \ref{app: Ach_Gap_UC_Re_3}.
\end{IEEEproof}
\section{Bidirectional Cooperation}\label{sec: BC}
Now, it is assumed that both of $C_{12}$ and $C_{21}$ can be non-zero. $C_{12}$ and $C_{21}$ are respectively used for the zero-forcing and relaying purposes as suggested by investigation of the unidirectional case.
Based on the observations obtained in the unidirectional cooperation case, achievable
schemes are proposed to characterize the capacity region of the GZIC with bidirectional cooperation up to 1.71 bits per user. Five scenarios are possible:
\renewcommand{\labelenumi}{\Alph{enumi}.}
\begin{enumerate}
    \item $a^2P_1\!\leq\!1$: It is explained that treating interference as noise makes ${\cal R}(0, 1)$ achievable. See Appendix \ref{app: Ach_Gap_BC_1} for details.
  \item $1 \leq a^2P_1\!\leq\!P_2\!+\!1$: In this case, since relaying can increase User 2's rate by one bit, $C_{21}$ is not used in this regime (similar to Section \ref{subsubsec: nc_Re}). Appendix \ref{app: Ach_Gap_BC_2} shows the zero-forcing technique utilized in Section \ref{subsec: ZF} is sufficient to achieve ${\cal R}(0.5, 1.5)$.
  \item $P_2+1 \leq a^2P_1, a^2 \leq 1$: Both relaying and zero-forcing techniques are used to show ${\cal R}(0.5,1.71)$ is achievable. To avoid the complexity of decoding four signals at Receiver 1, a compressed version of User 1's private signal is zero-forced at Receiver 2.
  \item $P_2+1 \leq a^2P_1, 1\leq a^2 \leq P_2+1$: In this regime, a simple combination of zero-forcing and relaying techniques proposed for the unidirectional case is used to prove ${\cal R}(1,1.21)$ is achievable.
  \item $P_2\!+\!1\!\le\!a^2P_1, P_2\!+\!1\!\leq\!a^2$: In this case, because Receiver 2 gets a very strong signal from
  Transmitter 1, there is not much benefit in using $C_{12}$. In fact, the same scheme employed for $C_{12}\!=\!0$ makes ${\cal R}(1,0.5)$ achievable.
  The gap analysis of Appendix \ref{app: Ach_Gap_UC_Re_3} is also valid for the case of $C_{12}>0$.
\end{enumerate}
It can be seen that for scenarios A, B, and E, the sum-capacity is approximated within 2 bits. In the following, scenarios C and D are further elucidated. Appendices \ref{app: Ach_Gap_BC_3}, and \ref{app: Ach_Gap_BC_4} argue the same gap holds for scenarios C and D, which assures the maximum gap of 2 bits on the sum-capacity for all regimes.
\setcounter{subsection}{2}
\subsection{$P_2+1 \leq a^2P_1, a \leq 1$}
A natural generalization of the schemes proposed for the unidirectional cooperation case, is
to consider the following signaling:
\begin{align*}
\begin{split}
X_1&=X_{1p}+X_{1c}+X_{1z}+X_{2r},\\
X_2&= X_{2p}-a X_{1z},
\end{split}
\end{align*}
where $X_{2r}$ is decoded at both receivers. To avoid the complexity of jointly decoding of four signals at Receiver 1,
User 1's private signal is compressed and sent to the other transmitter via the cooperative link with capacity $C_{12}$.\footnote{Instead of the compression, one might sequentially decode $(X_{1c}, X_{1z}, X_{2r})$
 and $X_{1p}$, similar to the approach of Section \ref{sec: PCCC}.} Then, Transmitter 2 zero-forces the compressed version of the signal. In particular, the following signaling is used:
\begin{align}
\begin{split}
X_1&=X_{1p}+X_{1c}+X_{2r},\\
X_2&= X_{2p}-\hat{X}_{1p},
\end{split}
\end{align}
where $a X_{1p}$ is compressed with distortion 1, \emph{i.e.}, $a X_{1p}\!=\!\hat{X}_{1p}\!+\!Z$,
with $Z\!\!\sim\!{\cal N}(0, 1)$. The compression, imposes the constraint $\C(a^2 P_{1p}\!-\!1)\!\le\!C_{12}$.
For the power allocation, it is recalled from the unidirectional cooperation case that at most half of Transmitter 2's power is allocated for zero-forcing to not harm
its own maximum rate by more than half a bit. Therefore, the following power allocation is considered:
\begin{align}
\begin{split}\label{eq: PA_Compression}
P_{1p}&=\frac{\min\{2^{2C_{12}},\frac{P_2}{2}+1\}}{a^2},\\
P_{1c}=P_{2r}&=\frac{P_1-P_{1p}}{2},\\
P_{2p}&=P_2-(a^2 P_{1p}-1).
\end{split}
\end{align}
The term containing $C_{12}$ is due to the compression rate constraint.
\begin{lemma}\label{lem: Ach_Gap_BC_3}
The region given in (\ref{eq: Rm1_1_c})-(\ref{eq: Rm1pR2_3_c}) is ${\cal R}(0.5,1.71)$ achievable.
\end{lemma}
\begin{IEEEproof}
See Appendix \ref{app: Ach_Gap_BC_3}.
\end{IEEEproof}
\subsection{$P_2+1 \leq a^2P_1, 1\leq a^2 \leq P_2+1$}
In this regime, the zero-forcing technique used for $C_{21}\!=\!0$, and the relaying technique used for $C_{12}\!=\!0$ are simply combined, \emph{i.e.},
\begin{align}
\begin{split}
X_1&=X_{1z}+X_{1c}+X_{2r},\\
X_2&= X_{2p}-a X_{1z}.
\end{split}
\end{align}
Similar to the unidirectional case, the following power allocation is used:
\begin{align*}
P_{1z}&=\frac{P_2}{2a^2},\\
P_{1c}=P_{2r}&=\frac{P_1-P_{1z}}{2},\\
P_{2p}&=\frac{P_2}{2}.
\end{align*}
\begin{lemma}\label{lem: Ach_Gap_BC_4}
The following region is ${\cal R}(1,1.21)$ achievable:
\begin{align}
R_1             &\le \C(\frac{P_1+\frac{P_2}{2a^2}}{2}) \label{eq: R1_1_bd_ag11f}\\
R_1             &\le \C(\frac{P_1-\frac{P_2}{2a^2}}{2})+C_{12} \label{eq: R1_2_bd_ag11f}\\
R_2             &\le \C(\frac{P_2}{2})+C_{21}\label{eq: R2_1_bd_ag11f}\\
R_2             &\le \C(\frac{2a^2P_1+P_2}{4})\label{eq: R2_2_bd_ag11f}\\
R_1+R_2         &\le \C(a^2P_1)+C_{12}-\frac{1}{2}\label{eq: R1pR2_1_bd_ag11f}\\
R_1+R_2         &\le \C(a^2P_1)+\C(\frac{P_2}{2a^2})-\frac{1}{2}\label{eq: R1pR2_2_bd_ag11f}\\
R_1+R_2         &\le \C(P_1)+\C(\frac{P_2}{2})-\frac{1}{2}.\label{eq: R1pR2_3_bd_ag11f}
\end{align}
\end{lemma}
\begin{IEEEproof}
See Appendix \ref{app: Ach_Gap_BC_4}.
\end{IEEEproof}
\section{Bidirectional v.s. Unidirectional Cooperation: A Numerical Analysis View}\label{sec: BCvsUC}
In the previous section, it has been observed that unidirectional cooperation is optimum in the constant gap sense for the following two cases: $1 \leq a^2P_1\!\leq\!P_2\!+\!1$, and $P_2\!+\!1\!\le\!\min\{a^2P_1, a^2\}$. In this section, however, various numerical results are presented to demonstrate that bidirectional
cooperation between the transmitters may provide better rate pairs compared to the unidirectional cooperation with
a relatively larger total cooperation capacity. To achieve this goal, it is assumed that the total cooperation capacity $C\triangleq C_{12}+C_{21}$ is fixed and can be arbitrarily distributed between the cooperative links.
The achievable region optimized over such distribution is compared to the capacity region or the outer bound region corresponding to the extreme case of unidirectional cooperation , \emph{i.e.}, the Cognitive Radio Z Channel (CRZC).\footnote{In the cognitive setup, the cognitive transmitter knows the whole message of the primary user \cite{MaricIT07}. This property can be modeled as having a GIC with unidirectional transmitter cooperation. In such a configuration, the capacity of the cooperative link suffices to be equal to the message rate of the primary user. This fact is described for the MAC with conferencing encoders in \cite{WillemsIT83}. In this section, it is shown that sharing that required capacity, or even smaller than that amount, between the bidirectional cooperative links can provide better rate pairs depending on the channel parameters.}

\begin{figure}[tb]
\centering
\subfigure[]{
\includegraphics[scale=0.39]{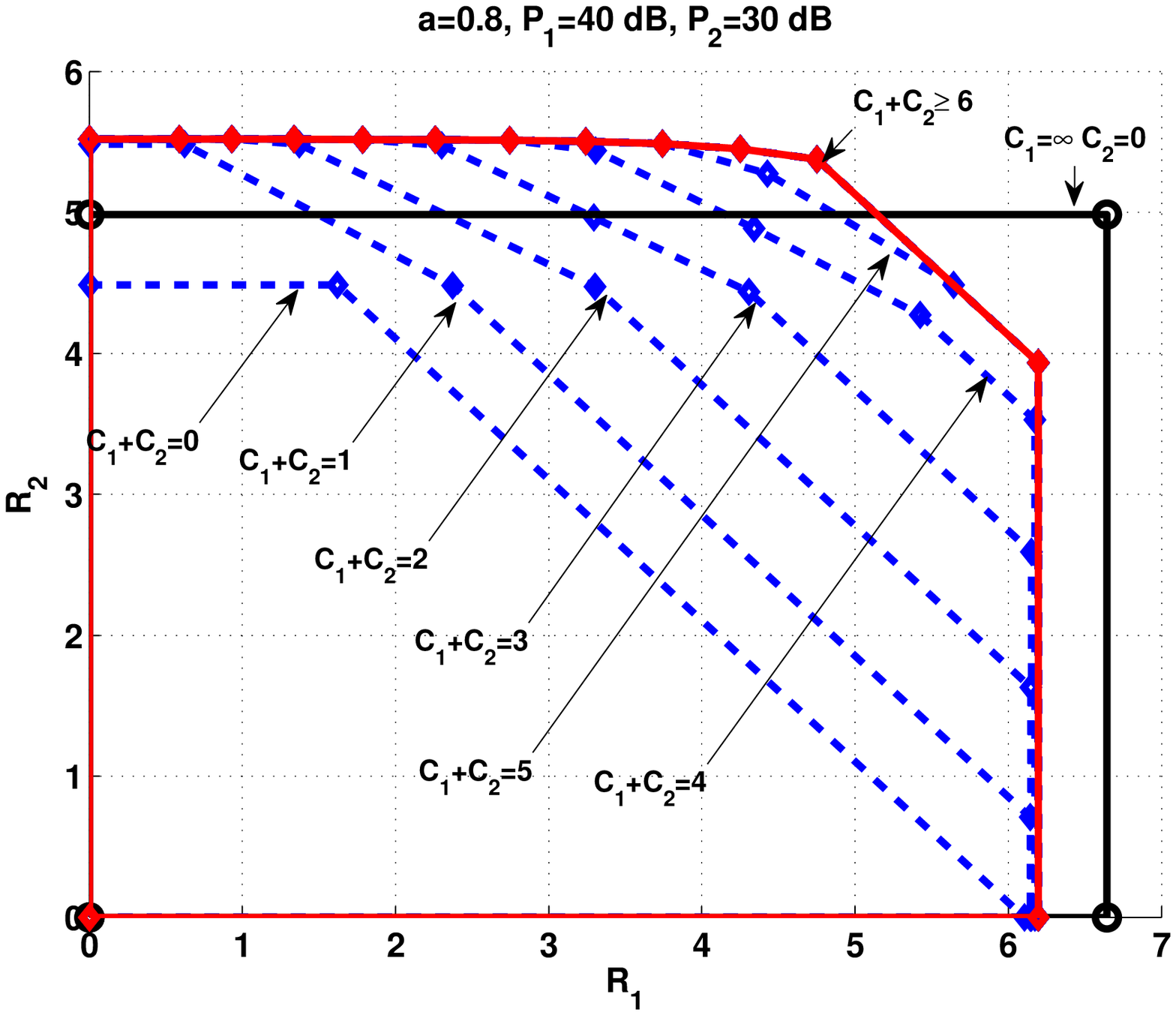}
\label{fig:subfig1}
}
\subfigure[]{
\includegraphics[scale=0.39]{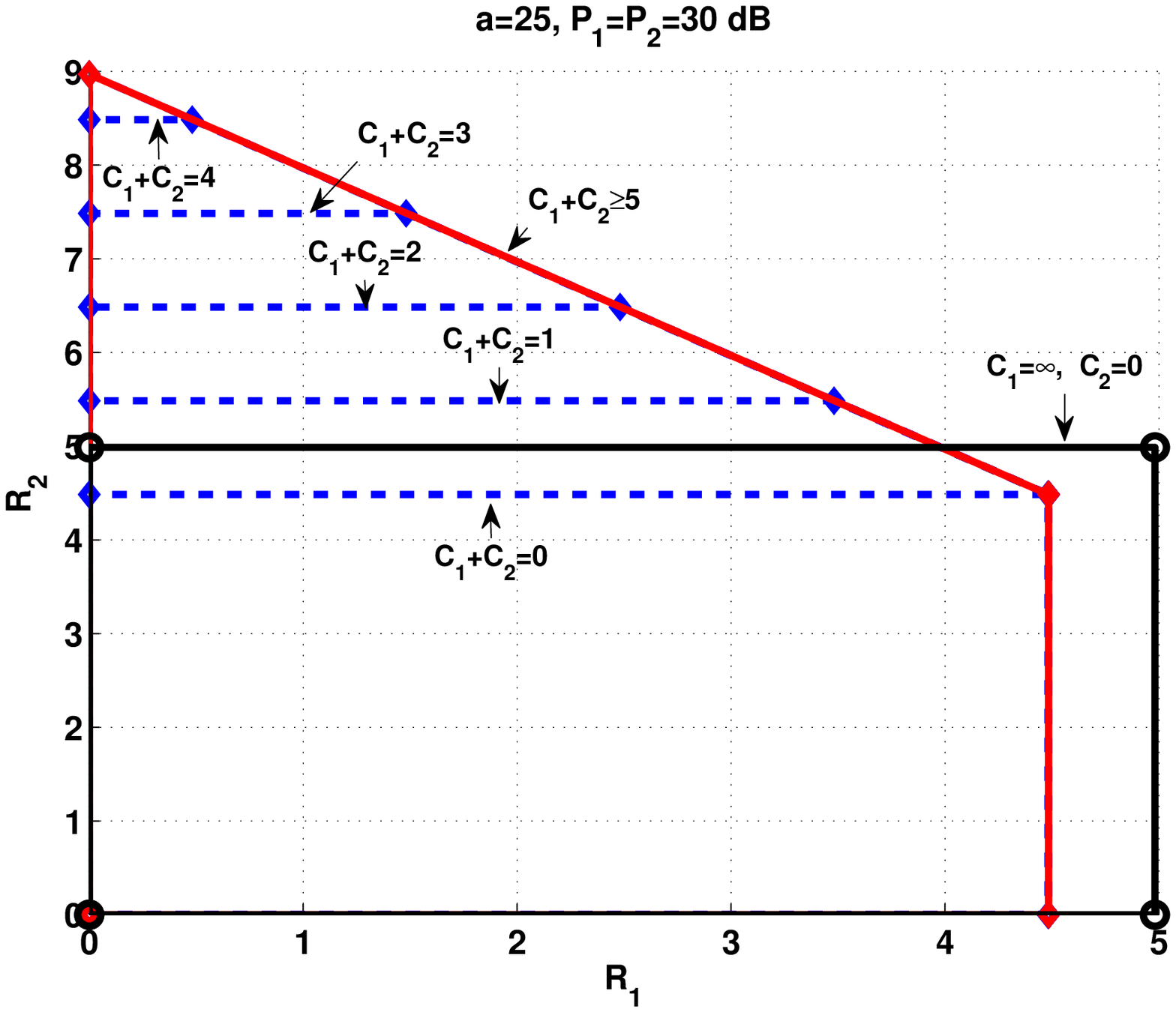}
\label{fig:subfig3}
}
\caption[Optional caption for list of figures]{Sharing the cooperative capacity between bidirectional cooperative links can enhance the maximum of $R_2$ compared to the type I CRZC. For $P_2+1\le a^2$ (not shown in this figure), the sum-rate can also be increased.}\label{fig:subfigureC1}
\end{figure}

\begin{figure}[tb]
\centering
\subfigure[]{
\includegraphics[scale=0.39]{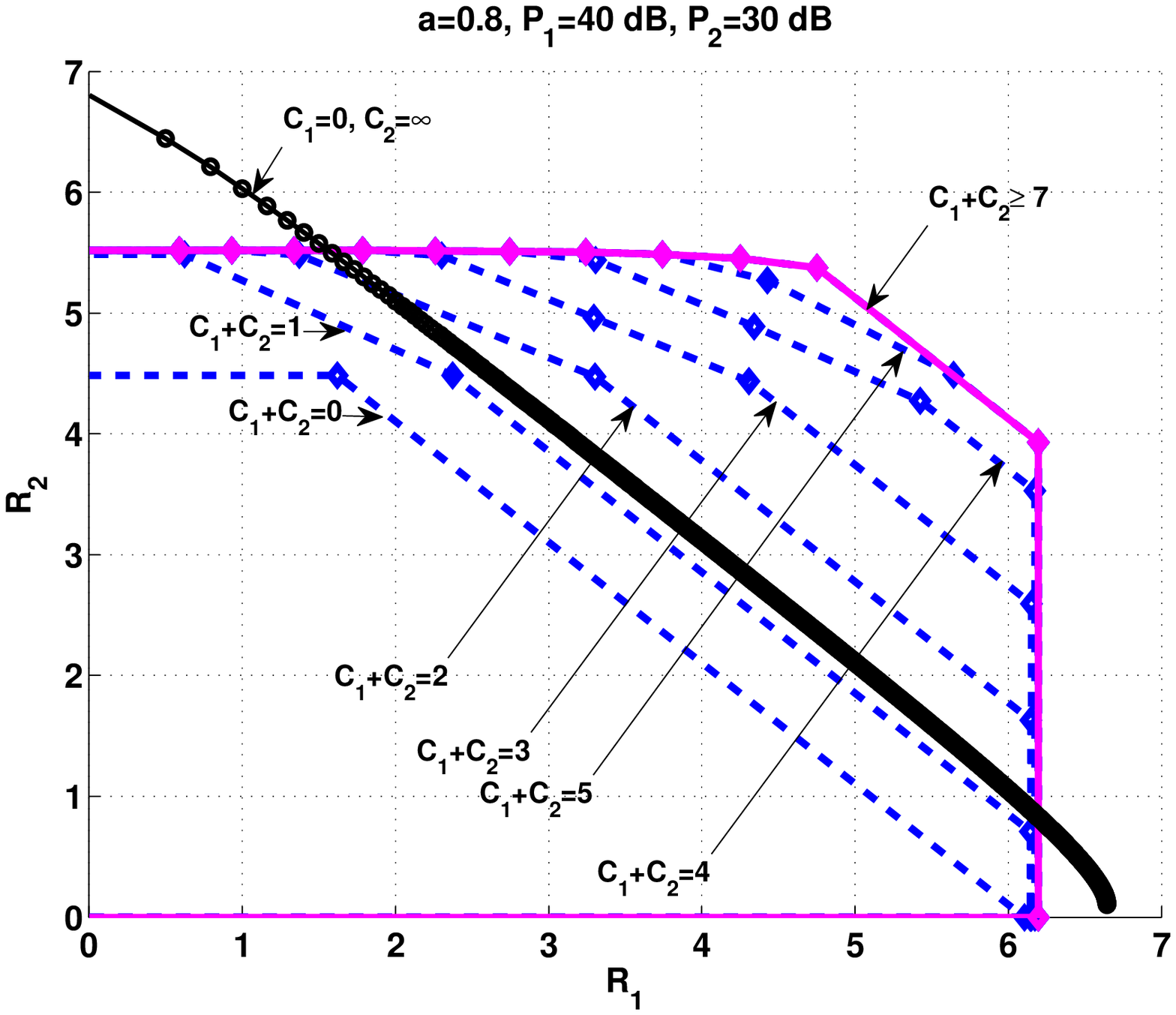}
\label{fig:subfig5}
}
\subfigure[]{
\includegraphics[scale=0.39]{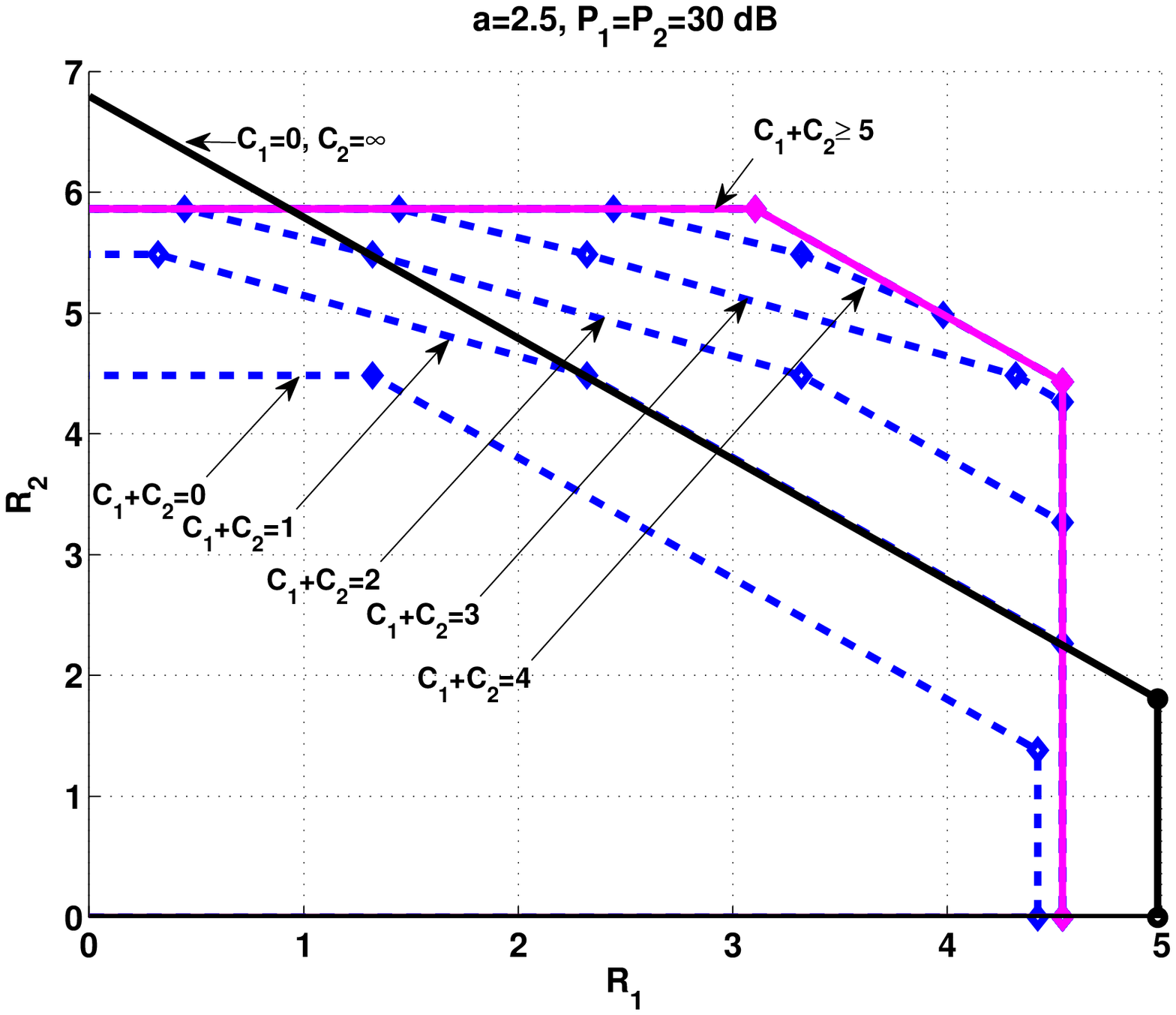}
\label{fig:subfig6}
}
\caption[Optional caption for list of figures]{Sharing the cooperative capacity between bidirectional cooperative links can improve the sum-rate compared to the type II CRZC.}\label{fig:subfigureC2}
\end{figure}

The CRZC can be either in the form of Fig. \ref{fig: ZF-RE model} (a) or Fig. \ref{fig: ZF-RE model} (b).
The former and the latter forms, respectively called type I and II, serve as the baseline for comparison in Figs. \ref{fig:subfigureC1} and \ref{fig:subfigureC2}, respectively. The capacity regions of the type I CRZC (for all $a$ values), and type II CRZC (for $a\le1$) are expressed by:
\begin{align*}
R_1^{\text{up}}&\le \C(P_1)\\
R_2^{\text{up}}&\le \C(P_2),
\end{align*}
and the union of the regions over $0\le \rho \le 1$
\begin{align*}
R_1^{\text{up}}&\le \C(\rho P_1)\\
R_2^{\text{up}}&\le \C\lp \frac{(\sqrt{a^2 (1-\rho) P_1}+\sqrt{P_2})^2}{1+a^2 \rho P_1}\rp,
\end{align*}
respectively \cite{WuIT07}. For type II CRZC with $1\!<\!a$, the upper bound region described by (\ref{eq: Ru1_1})-(\ref{eq: Ru1pRu2_2}) with $C_{12}\!=\!0$, and $C_{21}\!=\!\infty$ is used.

Figs. \ref{fig:subfigureC1} and \ref{fig:subfigureC2} evaluate the achievable rates for different values of the channel parameters.\footnote{To make figures more readable, $C_1$, and $C_2$ are used instead of $C_{12}$, and $C_{21}$, respectively.}
Looking at Fig. \ref{fig:subfigureC1}, it is seen that sharing the cooperative capacity can significantly increase the maximum of $R_2$, with respect to the type I CRZC, as $a$ gets larger. In contrast, Fig. \ref{fig:subfigureC2} shows
bidirectional cooperation with relatively smaller total cooperation capacity can substantially enhance the sum-rate compared to
 the type II CRZC.
\section{Conclusion}\label{sec: Conclusion}
It has been shown that for GZIC with bidirectional cooperation, ${\cal R}(1,1.71)$ is achievable. As another outcome of the gap analysis, the sum-capacity of the channel has been determined up to 2 bits. To obtain the results, basic communication techniques including Han-Kobayashi, zero-forcing, simple relaying, and transmission at the noise level schemes are employed. In addition, with the aid of signal compression or sequential decoding methods, the decoding complexity of the utilized schemes is limited to jointly decoding of at most three independent signals at each receiver.
It has been observed that unidirectional and bidirectional cooperation almost similarly perform (in the promised constant gap sense) in two scenarios:
\renewcommand{\labelenumi}{\arabic{enumi})}
\begin{enumerate}
  \item $C_{21}$ is not required when the relaying power is smaller than the direct transmission power, \emph{i.e.}, $a^2P_1\le P_2+1$.
  \item $C_{12}$ is not necessary when the relaying power and the relay gain are sufficiently large, \emph{i.e.}, $P_2+1 \le a^2P_1$, and $P_2+1 \le a^2$.
\end{enumerate}
Furthermore, it has been shown that properly sharing the total cooperative capacity between the bidirectional links can enhance the achievable rate pairs in some scenarios.
\appendices
\section{Achievable Rate and Gap Analysis for Zero-Forcing Scenario} \label{app: Gap_UC_ZF} 
The resulting rate constraints from error analysis at Decoder 1 are:
\vspace{-2pt}
\begin{align*}
R_{1p}&\leq \C(P_{1p})\\
R_{1z}&\leq \min\left\{\C(P_{1z}),C_{12}\right\}\\
R_{1c}&\leq \C(P_{1c})\\
R_{1p}+R_{1z}&\leq \C\lp P_{1p}+P_{1z}\rp\\
R_{1p}+R_{1c}&\leq \C(P_{1p}+P_{1c})\\
R_{1z}+R_{1c}&\leq \C(P_{1z}+P_{1c})\\
R_{1p}+R_{1z}+R_{1c}&\leq \C(P_1),
\end{align*}
and the constraints for Decoder 2 are:
\vspace{-3pt}
\begin{align*}
R_{2p}&\leq \C(\frac{P_{2p}}{d})\\
R_{2p}+R_{1c}&\leq \C(\frac{a^2P_{1c}+P_{2p}}{d}),
\end{align*}
where $d$ is defined in (\ref{eq: d}).
Note that there is no individual rate constraint on $R_{1c}$ at Decoder 2 because the joint typical decoder does not declare an error in the case that only $X_{1c}$ is wrongly decoded \cite{Motani_IT08}.
Fourier-Motzkin Elimination (FME) \cite{book:LP} is applied to obtain the constraints in terms of $R_1\!\triangleq\!R_{1p}\!+\!R_{1z}\!+\!R_{1c}$, and $R_2\!\triangleq\!R_{2p}$. Removing the redundant inequalities due to the polymatroid structure of the rate constraints at each decoder, leads to the region (\ref{eq: Ach_ZIC_UC_ZF_R1})-(\ref{eq: Ach_ZIC_UC_ZF_R1pR2}). To simplify the gap analysis, it is noted that the subsequent region is also achievable:
\vspace{-1pt}
\begin{align}
R_1&\leq \C(\frac{P_{1}}{2})\label{eq: ach_zu_zf_s_R1}\\
R_{2}&\leq \C(\frac{P_2}{4})\label{eq: ach_zu_zf_s_R2}\\
R_{1}+R_{2}&\leq \C(\Ppr)+\C(\frac{a^2P_1-1+P_2}{4})+C_{12}\label{eq: ach_zu_zf_s_R1pR2_1}\\
R_{1}+R_{2}&\leq \C(\frac{\min\{a^2P_1-1, P_2\}+2}{2a^2})+\C(\frac{\max\{a^2P_1-1, P_2\}}{4}).\label{eq: ach_zu_zf_s_R1pR2_2}
\end{align}
It is clear that the preceding region is smaller than that of (\ref{eq: Ach_ZIC_UC_ZF_R1})-(\ref{eq: Ach_ZIC_UC_ZF_R1pR2}).
To obtain this region, we use (\ref{eq: P1z}) and set $P_{1c}=\frac{P_1-\Ppr}{2}$ and $P_{2p}=\frac{P_2}{2}$ in the 
achievable region (\ref{eq: Ach_ZIC_UC_ZF_R1})-(\ref{eq: Ach_ZIC_UC_ZF_R1pR2}). 
In addition, in getting (\ref{eq: ach_zu_zf_s_R1}) and (\ref{eq: ach_zu_zf_s_R1pR2_2}) 
we notice $\frac{P_1}{2}\le P_{1p}+P_{1c}$ and $\max\{a^2P_{1c}, P_{2p}\}\le a^2P_{1c}+P_{2p}$. In the following, the region is compared to the upper bound to show that ${\cal R}(0.5, 1)$ is achievable:
\begin{align*}
(\ref{eq: Ru1_1})-(\ref{eq: ach_zu_zf_s_R1})&\le 0.5\\
(\ref{eq: Ru2_1})-(\ref{eq: ach_zu_zf_s_R2})&\le 1\\
(\ref{eq: Ru1pRu2_1})-(\ref{eq: ach_zu_zf_s_R1pR2_1})&\stackrel{(a)}{\le} 1.5\\
(\ref{eq: Ru1_1})+(\ref{eq: Ru2_1})-(\ref{eq: ach_zu_zf_s_R1pR2_2})&\le 1.5.
\end{align*}
By $(i)\pm(j)$, we mean the sum/difference of the right hand side of Eqs. (i) and (j). To attain $(a)$, the fact that
$\C(\frac{\max\{1-a^2, 0\}P_1}{1+a^2P_1})\!\le\!\C(\frac{1}{a^2})$ is used. It is remarked that $R_{1p}\!<\!1$ for the case
 of $1\!<\!a$. Therefore, we can set $P_{1p}\!=\!0$, \emph{i.e.}, we do not use $X_{1p}$ in order to simplify the scheme. It can be shown that ${\cal R}(0.5,0.5)$ is achievable in this case.
\section{Achievable Rate and Gap Analysis for Relaying Scenario}
In this appendix, the achievable rate and gap analysis corresponding to the three scenarios identified in Section \ref{subsec: Re} are derived in detail.
\subsection{Non-Cooperative Case}\label{app: Ach_Gap_UC_Re_1}
The decoding rules lead to the following rate constraints at Decoder 1 and Decoder 2:

Decoder 1:
\begin{align*}
R_{1p}&\leq \C(P_{1p})\\
R_{1c}&\leq \C(P_{1c})\\
R_{1p}+R_{1c}&\leq \C(P_1),
\end{align*}

Decoder 2:
\begin{align*}
R_{2p}&\leq \C(\frac{P_2}{2})\\
R_{2p}+R_{1c}&\leq \C(\frac{P_2+a^2P_{1c}}{2}).
\end{align*}

Noting that $a^2P_1-1 \le P_2$, it is straightforward to prove the region (\ref{eq: R1_Re_nc})-(\ref{eq: R1pR2_Re_nc}), which is obtained by applying FME, is ${\cal{R}}(0,1.5)$ achievable.

To find the worst case gap, (\ref{eq: R1_Re_nc}), (\ref{eq: R2_Re_nc}), and (\ref{eq: R1pR2_Re_nc}) are respectively compared to the upper bounds (\ref{eq: Ru1_1}), (\ref{eq: Ru2_2}), and (\ref{eq: Ru1pRu2_1}). As in the conventional GIC for $1\!<\!a$, the private signal is not needed, and therefore, ${\cal{R}}(0,1)$ is achievable.

\subsection{Common Cooperative Case (Proof of Lemma \ref{lem: Ach_Gap_UC_Re_2})} \label{app: Ach_Gap_UC_Re_2}
The decoding rules impose the following constraints:

Decoder 1:

\begin{align}
R_{1p}&\le \C(P_{1p})\label{eq: R1p_Dec1}\\
R_{1c}&\le \C(P_{1c})\\
R_{1p}+R_{1c}&\le \C(P_{1p}+P_{1c})\\
R_{1p}+R_{2r}&\le \C(P_{1p}+P_{2r})\\
R_{1c}+R_{2r}&\le \C(P_{1c}+P_{2r})\label{eq: R1cpR2r_Dec1}\\
R_{1p}+R_{1c}+R_{2r}&\le \C(P_{1})\label{eq: R1ppR1cpR2r_Dec1},
\end{align}

Decoder 2:

\begin{align}
R_{2p}&\le \C(\frac{P_{2p}}{d})\label{eq: R2p_Dec2}\\
R_{2r}&\le \min\{C_{21},\C(\frac{a^2 P_{2r}}{d})\}\\
R_{2p}+R_{2r}&\le \C(\frac{a^2 P_{2r}+P_{2p}}{d})\\
R_{2p}+R_{1c}&\le \C(\frac{P_{2p}+a^2 P_{1c}}{d})\\
R_{2r}+R_{1c}&\le \C(\frac{a^2 (P_{2r}+P_{1c})}{d})\label{eq: R1cpR2r_Dec2}\\
R_{2p}+R_{2r}+R_{1c}&\le \C(\frac{P_{2p}+a^2 (P_{2r}+P_{1c})}{d}),\label{eq: R2ppR2rpR1cp_Dec2}
\end{align}
where $d$ is defined in (\ref{eq: d}).

Defining $R_1\triangleq R_{1p}+R_{1c}$ and $R_2\triangleq R_{2p}+R_{2r}$, applying FME, and removing redundant inequalities, lead to the following rate constraints:

\begin{align}
R_{1}&\le \C(P_{1p}+P_{1c})\label{eq: R1_1_c}\\
R_{1}&\le \C(P_{1p})+\C(\frac{a^2 (P_{2r}+P_{1c})}{d})\label{eq: R1_2_c}\\
R_{2}&\le \C(\frac{P_{2p}}{d})+C_{21}\label{eq: R2_1_c}\\
R_{2}&\le \C(\frac{a^2 P_{2r}+P_{2p}}{d})\label{eq: R2_2_c}\\
R_{1}+R_{2}&\le \C(P_{1})+\C(\frac{P_{2p}}{d})\label{eq: R1pR2_1_c}\\
R_{1}+R_{2}&\le \C(\frac{P_{2p}+a^2 P_{1c}}{d})+\C(P_{1p}+P_{2r})\label{eq: R1pR2_2_c}\\
R_{1}+R_{2}&\le \C(\frac{P_{2p}+a^2 (P_{2r}+P_{1c})}{d})+\C(P_{1p})\label{eq: R1pR2_3_c}\\
R_{1}+R_{2}&\le \C(\frac{P_{2p}+a^2 P_{1c}}{d})+\C(P_{1p})+C_{21}\label{eq: R1pR2_4_c}\\
2R_1+R_2&\le \C(\frac{P_{2p}+a^2 P_{1c}}{d})+\C(\frac{a^2 (P_{2r}+P_{1c})}{d})+2\C(P_{1p})\label{eq: 2R1pR2_1_c}\\
2R_1+R_2&\le \C(\frac{P_{2p}+a^2 P_{1c}}{d})+\C(P_{1p})+\C(P_{1})\label{eq: 2R1pR2_2_c}\\
R_1+2R_2&\le \C(\frac{P_{2p}}{d})+\C(\frac{P_{2p}+a^2 (P_{2r}+P_{1c})}{d})+\C(P_{1p}+P_{2r})\label{eq: R1p2R2_1_c}\\
R_1+2R_2&\le 2\C(\frac{P_{2p}}{d})+\C(P_{2r}+P_{1c})+\C(P_{1p}+P_{2r}).\label{eq: R1p2R2_2_c}
\end{align}
\vspace{-3pt}
It is noted that (\ref{eq: R1p2R2_2_c}) is redundant since $(\ref{eq: R1p2R2_1_c})\le (\ref{eq: R1p2R2_2_c})$ as verified below for two cases of $a\le 1$, and $1<a$:
\begin{align*}
(\ref{eq: R1p2R2_2_c})-(\ref{eq: R1p2R2_1_c})&=(\ref{eq: R2p_Dec2})+(\ref{eq: R1cpR2r_Dec1})-(\ref{eq: R2ppR2rpR1cp_Dec2})\\
{} &\ge (\ref{eq: R2p_Dec2})+(\ref{eq: R1cpR2r_Dec2})-(\ref{eq: R2ppR2rpR1cp_Dec2})\ge 0 \quad\ \ \!\quad\quad\quad\quad \text{for} \ a\le 1\\
{} &=\C(P_2)\!+\!\C(P_1)\!-\!\C(P_2\!+\!a^2P_1)\stackrel{(\star)}{\ge} 0 \quad \text{for} \ 1< a,
\end{align*}
where to prove $(\star)$, $a^2$ is replaced by its maximum value, \emph{i.e.}, $P_2+1$.

We set $P_{2r}=P_{1c}=\frac{P_1-P_{1p}}{d}$. This power allocation policy not only does not decrease the maximum rates of both users by more than 0.5 bit, but also can simplify the achievable region by making some inequalities redundant. In particular, if we decrease (\ref{eq: R1pR2_3_c}) by 0.5 bit, then the region described by (\ref{eq: Rm1_1_c})-(\ref{eq: Rm1pR2_3_c}) is achievable.
This is due to:
\begin{align*}
(\ref{eq: R1_1_c})&\le (\ref{eq: R1_2_c})\\
(\ref{eq: R1pR2_3_c})-0.5 & \stackrel{(a)}{\le} (\ref{eq: R1pR2_1_c})\\
(\ref{eq: R1pR2_3_c})-0.5 & \le \min\{(\ref{eq: R1pR2_2_c}), (\ref{eq: R1pR2_4_c})\}\\
(\ref{eq: R1_2_c})+(\ref{eq: R1pR2_3_c})-0.5 & \le (\ref{eq: 2R1pR2_1_c})\\
(\ref{eq: R1_1_c})+(\ref{eq: R1pR2_3_c})-0.5 & \le (\ref{eq: 2R1pR2_2_c})\\
(\ref{eq: R2_2_c})+(\ref{eq: R1pR2_3_c})-0.5& \stackrel{(b)}{\le} (\ref{eq: R1p2R2_1_c}).
\end{align*}
To establish $(a)$ and $(b)$, we use the fact that $P_2+1 \le a^2 P_1$. The following steps are proceeded to prove the region is ${\cal R}(0.5,1.5)$ achievable for $a\le 1$:
\begin{align*}
(\ref{eq: Ru1_1})-(\ref{eq: Rm1_1_c})&\leq 0.5\\
(\ref{eq: Ru2_1})-(\ref{eq: Rm2_1_c})&\leq 0.5\\
(\ref{eq: Ru2_2})-(\ref{eq: Rm2_2_c})&\leq 1.5\\
(\ref{eq: Ru1pRu2_1})-(\ref{eq: Rm1pR2_3_c})&\leq 1.5,
\end{align*}
and ${\cal R}(0.5,0.5)$ achievable for $1<a$:
\begin{align*}
(\ref{eq: Ru1_1})-(\ref{eq: Rm1_1_c})&\leq 0.5\\
(\ref{eq: Ru2_1})-(\ref{eq: Rm2_1_c})&= 0\\
(\ref{eq: Ru1pRu2_1})-(\ref{eq: Rm1pR2_3_c})&\leq 1.
\end{align*}
It is seen that the worst gap for this case, which is due to (\ref{eq: Rm2_2_c}) or (\ref{eq: Rm1pR2_3_c}), can be further reduced by sending $X_{2r}$ from both transmitters. To achieve the smaller gap goal, the term $2a^2P_1+2P_2$ in the upper bounds (\ref{eq: Ru2_2}) and (\ref{eq: Ru1pRu2_1}) can also be tightened to $(a\sqrt{P_1}+\sqrt{P_2})^2$.
\subsection{Private-Common Cooperative Case (Proof of Lemma \ref{lem: Ach_Gap_UC_Re_3})} \label{app: Ach_Gap_UC_Re_3}
The decoding rules enforce the following rate constraints:

Decoder 1:
\begin{align*}
R_{1c}&\le \C(\frac{P_{1c}}{2})\\
R_{1c}+R_{2r_c}&\le \C(\frac{P_1-1}{2}).
\end{align*}
The factor $2$ in the denominators is because of treating $X_{2r_p}$ as noise.

Decoder 2:
\begin{align*}
R_{2r_p}&\le C'\\
R_{2r_c}&\le \min\{C_{21}-C', \C(\frac{a^2 P_{2r_c}}{P_2+1})\}\\
R_{2r_p}+R_{2r_c}&\le \C(\frac{a^2(P_{2r_c}+1)}{P_2+1})\\
R_{2r_p}+R_{1c}&\le \C(\frac{a^2(P_{1c}+1)}{P_2+1})\\
R_{2r_c}+R_{1c}&\le \C(\frac{a^2(P_1-1)}{P_2+1})\\
R_{2r_p}+R_{2r_c}+R_{1c}&\le \C(\frac{a^2P_1}{P_2+1})\\
R_{2p}&\le \C(P_2),
\end{align*}
where $C'=\min\{\C(\frac{a^2}{P_2+1}),C_{21}\}$. The expression $P_2+1$ in denominators comes from the sequential decoding of $(X_{2r_p}, X_{2r_c}, X_{1c})$, and $X_{2p}$. The constraint on $R_{2p}$ is obtained due to the assumption that $X_{2p}$ is decoded after $(X_{2r_p}, X_{2r_c}, X_{1c})$ are decoded and their effect is subtracted from the received signal. FME is used to rewrite the constraints in terms of $R_1\triangleq R_{1c}$, and $R_2\triangleq R_{2p}+R_{2r_p}+R_{2r_c}$. After removing inactive inequalities according to the operating regime, \emph{i.e.}, $P_2+1\le a^2$, we attain:
\begin{align*}
R_1&\le \C(\frac{P_{1c}}{2})\\
R_2&\le \C(P_2)+C_{21}\\
R_2&\le \C\lp a^2 (P_{2r_c}+1)+P_2 \rp\\
R_1+R_2&\le \C(\frac{P_1-1}{2})+C'+\C(P_2)\\
R_1+R_2&\le \C(a^2 P_1+P_2)\\
R_1+R_2&\le \C\lp a^2 (P_{1c}+1)+P_2 \rp+C_{21}-C'\\
R_1+R_2&\le \C\lp a^2 (P_{1c}+1)+P_2 \rp+\C(\frac{a^2 P_{2r_c}}{P_2+1})\\
2R_1+R_2&\le \C\lp a^2 (P_{1c}+1)+P_2 \rp+\C(\frac{P_1-1}{2}).
\end{align*}
Substituting the allocated powers provides:
\begin{align}
R_1&\le \C(\frac{P_1-1}{4})\label{eq: R1_Re_1}\\
R_2&\le \C(P_2)+C_{21}\label{eq: R2_Re_1}\\
R_2&\le \C(\frac{a^2 (P_1+1)}{2}+P_2)\label{eq: R2_Re_2}\\
R_1+R_2&\le \C(\frac{P_1-1}{2})+C'+\C(P_2)\label{eq: R1pR2_Re_1}\\
R_1+R_2&\le \C(a^2 P_1+P_2)\label{eq: R1pR2_Re_2}\\
R_1+R_2&\le \C(\frac{a^2(P_1+1)}{2}+P_2)+C_{21}-C'\label{eq: R1pR2_Re_3}\\
R_1+R_2&\le \C(\frac{a^2(P_1+1)}{2}+P_2)+\C(\frac{a^2 P_1-1}{2(P_2+1)})\label{eq: R1pR2_Re_4}\\
2R_1+R_2&\le \C(\frac{a^2(P_1+1)}{2}+P_2)+\C(\frac{P_1-1}{2}).\label{eq: 2R1pR2_Re_1}
\end{align}
If we deduct 0.5 bit from (\ref{eq: R1pR2_Re_2}), some of the inequalities become redundant, since:
\begin{align*}
(\ref{eq: R1pR2_Re_2})-0.5\le (\ref{eq: R2_Re_2}), (\ref{eq: R1pR2_Re_3}), (\ref{eq: R1pR2_Re_4}),\\
\min\{(\ref{eq: R1_Re_1})+(\ref{eq: R2_Re_1}),(\ref{eq: R1pR2_Re_2})-0.5\}\le (\ref{eq: R1pR2_Re_1}),\\
(\ref{eq: R1_Re_1})+(\ref{eq: R1pR2_Re_2})-0.5\le (\ref{eq: 2R1pR2_Re_1}),
\end{align*}
leading to the achievable region (\ref{eq: Ach_Re_3}).
Now, the simplified region (\ref{eq: Ach_Re_3}) is compared to the upper bounds to show it is ${\cal R}(1,0.5)$ achievable:
\begin{align*}
\delta_{R_1}&\le 1\\
\delta_{R_2}&= 0\\
\delta_{R_1+R_2}&\le \C(5).
\end{align*}
Because $P_2+1 \le a^2$ in this regime, the upper bound (\ref{eq: Ru1pRu2_2}) is enlarged to $\C(3a^2P_1+P_1+2P_2)$ to be used in proving the gap on the sum-rate.
\section{Gap Analysis for Bidirectional Cooperation}\label{app: Ach_Gap_BC}
Most of the detailed gap analysis for the bidirectional cooperation case is provided below.
\subsection{$a^2P_1\leq 1$}\label{app: Ach_Gap_BC_1}
\vspace{-2pt}
Treating interference as noise and not using the cooperative links lead to the following ${\cal R}(0, 1)$ achievable region:
\begin{align}
R_1&\le \C(P_1) \label{eq: R1_1_BC_S1}\\
R_2&\le \C(\frac{P_2}{a^2 P_1+1}).\label{eq: R2_1_BC_S1}
\end{align}
To prove the gap, we note that
\begin{align*}
(\ref{eq: Ru1_1})-(\ref{eq: R1_1_BC_S1})&=0\\
(\ref{eq: Ru2_2})-(\ref{eq: R2_1_BC_S1})&\le \C(2a^2P_1+2P_2)-\C(\frac{a^2P_1+P_2-1}{2})\le 1.
\end{align*}
\subsection{$1 \leq a^2P_1\!\leq\!P_2\!+\!1$}\label{app: Ach_Gap_BC_2}
In this regime, the achievable scheme based on the zero-forcing technique, used for unidirectional case, is shown to be ${\cal R}(0.5, 1.5)$ achievable for the bidirectional case. In Appendix \ref{app: Gap_UC_ZF}, the region (\ref{eq: ach_zu_zf_s_R1})-(\ref{eq: ach_zu_zf_s_R1pR2_2}) is proposed to simplify the gap analysis, and also to prove ${\cal R}(0.5, 1)$ is achievable for the case of $C_{21}=0$. When $0 < C_{21}$, we modify the achievable region as well as the gap analysis to show ${\cal R}(0.5, 1.5)$ is achievable. It can be readily shown that the region is still achievable if we replace $R_2\le \C(\frac{P_2}{2})$ in (\ref{eq: ach_zu_zf_s_R2}) by $R_2\le \C(P_2-\frac{a^2P_1-1}{2})$. We remark that both regions considered in this appendix and Appendix \ref{app: Gap_UC_ZF} are inside the achievable region given in (\ref{eq: Ach_ZIC_UC_ZF_R1})-(\ref{eq: Ach_ZIC_UC_ZF_R1pR2}). We compare the upper bound to the modified region to
show ${\cal R}(0.5, 1.5)$ is achievable:
\begin{align*}
(\ref{eq: Ru1_1})-(\ref{eq: ach_zu_zf_s_R1})&\le 0.5\\
(\ref{eq: Ru2_2})-\C(P_2-\frac{a^2P_1-1}{2})&\stackrel{(a)}{\le}1.5\\
(\ref{eq: Ru1pRu2_1})-(\ref{eq: ach_zu_zf_s_R1pR2_1})&\le 1.5\\
(\ref{eq: Ru1pRu2_2})-(\ref{eq: ach_zu_zf_s_R1pR2_2})&\le \C(P_1P_2+P_1+4P_2+2)-(\ref{eq: ach_zu_zf_s_R1pR2_2}) \le 2.
\end{align*}
Here, we provide the proof for $(a)$. First we define some parameters:
\begin{align*}
x&\triangleq a^2P_1-1\\
v&\triangleq\frac{2x+2P_2+3}{-\frac{x}{2}+P_2+1}\\
u&\triangleq \C(2a^2P_1+2P_2)-\C(P_2-\frac{a^2P_1-1}{2})\\
{}&=\frac{1}{2}\log (v).
\end{align*}
It is easy to see that the derivative of $v$ with respect to $x$ is always positive. Therefore, the maximum value of $x$ provides the maximum value of $v$, and consequently $u$.
Hence, it is straightforward to show
\begin{align*}
\max\limits_{0\le x \le P_2}\{v\}&= \frac{4P_2+3}{\frac{P_2}{2}+1}\le 8,
\end{align*}
which proves $(a)$.
\subsection{$P_2+1 \leq a^2P_1, a \leq 1$}\label{app: Ach_Gap_BC_3}
The received signals for this signaling are:
\begin{align*}
Y_1&=X_{1p}+X_{1c}+X_{2r}+N_1,\\
Y_2&=X_{2p}+a X_{1c}+a X_{2r}+ Z+ N_2,
\end{align*}
where $N_1, N_2, \text{and} \ Z \ (\text{the compression noise}) \sim {\cal N}(0, 1)$.
The decoding rules impose the same constraints as (\ref{eq: R1p_Dec1})-(\ref{eq: R2ppR2rpR1cp_Dec2}) with $P_{2p}\!=\!P_2\!-\!(a^2P_{1p}\!-\!1)$ instead of $P_{2p}\!=\!P_2$ (due to zero-forcing). Therefore, FME provides the rate region given in (\ref{eq: R1_1_c})-(\ref{eq: R1p2R2_1_c}). Here, it is shown that the region (\ref{eq: Rm1_1_c})-(\ref{eq: Rm1pR2_3_c}) is achievable. First, it is noted that similar to Appendix \ref{app: Ach_Gap_UC_Re_2}, (\ref{eq: R1pR2_2_c}), and (\ref{eq: R1pR2_4_c})-(\ref{eq: 2R1pR2_2_c}) are redundant. In addition, the power allocation policy (\ref{eq: PA_Compression}) makes
\vspace{-3pt}
\begin{align*}
(\ref{eq: R1_1_c})&\stackrel{(a)}{\le} (\ref{eq: R1_2_c})\\
(\ref{eq: R1pR2_3_c})-0.5 & \stackrel{(b)}{\le} (\ref{eq: R1pR2_1_c})\\
(\ref{eq: R2_2_c})+(\ref{eq: R1pR2_3_c})-0.5& \stackrel{(c)}{\le} (\ref{eq: R1p2R2_1_c}),
\end{align*}
since
\renewcommand{\labelenumi}{\emph{(\alph{enumi})}.}
 \begin{enumerate}
 \item
 \begin{align*}
 \C(P_{1p})+\C(\frac{a^2(P_1-P_{1p})}{2})&\stackrel{(\circ)}{\ge} \C(P_{1p}+\frac{a^2(P_1-P_{1p})}{2}+\frac{P_1-P_{1p}}{2})\\
 {}&\ge \C(\frac{P_1+P_{1p}}{2}),
 \end{align*}
 \item
 \begin{align*}
 (\ref{eq: R2_2_c})+(\ref{eq: R1pR2_3_c})-0.5&\stackrel{(\star)}{\le} \C(\frac{P_{2p}+a^2(P_1-P_{1p})}{2}+P_{1p}+\frac{P_{1p}P_{2p}}{2}+\frac{(\frac{P_2}{2}+1)(P_1-P_{1p})}{2})-0.5\\
 {}&\le \C(P_1+\frac{P_{2p}}{2}+\frac{P_1P_2}{4})\\
 {}&\stackrel{(\diamond)}{\le} (\ref{eq: R1pR2_1_c}),
 \end{align*}
 \item
 \begin{align*}
\C(P_{1p})+\C(\frac{a^2P_{2r}+P_{2p}}{2})&\stackrel{(\star)}{\le} \C(P_{1p}+\frac{a^2P_{2r}+P_{2p}}{2}+\frac{\frac{P_2}{2}+1}{2}P_{2r}+\frac{P_{1p}P_{2p}}{2})\\
{}&\stackrel{(\diamond)}{\le}\C(P_{1p}+\frac{P_{2r}+P_{2p}}{2}+\frac{P_{2p}+1}{2}P_{2r}+\frac{P_{1p}P_{2p}}{2})\\
{}&\le 0.5+\C(\frac{P_{2p}}{2})+\C(P_{1p}+P_{2r}),
 \end{align*}
 where $(\circ)$, $(\star)$, and $(\diamond)$ are correct due to $1 \le a^2P_{1p}$, $a^2 P_{1p}\le\frac{P_2}{2}+1$, and $\frac{P_2}{2}\le P_{2p}$, respectively. It is remarked that the above proofs are also valid for the case of $a\le 1$ in Appendix \ref{app: Ach_Gap_UC_Re_2}.
 \end{enumerate}
To analyze the gap, we note that $\frac{P_2}{2}\leq P_{2p}$ assures
\begin{align*}
(\ref{eq: Ru1_1})-(\ref{eq: Rm1_1_c})&\leq 0.5\\
(\ref{eq: Ru2_1})-(\ref{eq: Rm2_1_c})&\leq 1\\
(\ref{eq: Ru2_2})-(\ref{eq: Rm2_2_c})&\stackrel{(\star)}{\le} \C(\frac{29}{3}),
\end{align*}
where $(\star)$ is true since $P_2\le a^2P_1-1$.

Now, if $P_{1p}=\frac{2^{C12}}{a^2}$, then adding both sides of the next three inequalities verifies that $(\ref{eq: Ru1pRu2_1})-(\ref{eq: Rm1pR2_3_c})\leq 1.5$.
\begin{align*}
\C(\frac{(1-a^2)P_1}{a^2P_1+1})+C_{12}-\C(\frac{2^{C12}}{a^2})&\le 0\\
\C(a^2 P_1)-\C(\frac{P_{2p}+a^2(P_1-P_{1p})}{2}) & \le 0.5\\
\C(2a^2P_1+2P_2)-\C(a^2 P_1) & \le 1.
\end{align*}

If $P_{1p}=\frac{P_2+2}{2 a^2}$, we have $P_{2p}=\frac{P_2}{2}$, and consequently,
\begin{align*}
\frac{P_{2p}+a^2(P_1-P_{1p})}{2}&=\frac{a^2P_1-1}{2}, \ \text{and}\\
(\ref{eq: Ru1pRu2_2})-(\ref{eq: Rm1pR2_3_c})&=\frac{1}{2}\log\lp\frac{1+P_1P_2+P_1(1+2a^2)+2P_2}{(1+\frac{P_2+2}{2a^2})(\frac{a^2P_1+1}{2})}\rp\\
{}&\le 1.5.
\end{align*}
Therefore, ${\cal R}(0.5,1.71)$ is achievable. It is also observed that $(\ref{eq: Rm1pR2_3_c})\le (\ref{eq: Rm1_1_c})+(\ref{eq: Rm2_2_c})$, which guarantees that the sum-rate is within 2 bits of the sum-capacity in this regime.
\subsection{$P_2+1 \leq a^2P_1, 1\leq a^2 \leq P_2+1$}\label{app: Ach_Gap_BC_4}
The decoding rules impose similar constraints as (\ref{eq: R1p_Dec1})-(\ref{eq: R2ppR2rpR1cp_Dec2}) with $R_{1p}\le \C(P_{1p})$ replaced by $R_{1z}\le \min\{C_{12},\C(P_{1z})\}$. FME is applied to write the constraints in the format of $R_1\triangleq R_{1z}+R_{1c}$, and $R_2\triangleq R_{2p}+R_{2r}$:
\begin{align}
R_1             &\le \C(P_{1z}+P_{1c})\label{eq: R1_1_bd_ag11}\\
R_1             &\le \C(P_{1c})+C_{12}\label{eq: R1_2_bd_ag11}\\
R_2             &\le \C(P_{2p})+C_{21}\label{eq: R2_2_bd_ag11}\\
R_2             &\le \C(P_{2p}+a^2 P_{2r})\label{eq: R2_1_bd_ag11}\\
R_2             &\le \C(P_{1c}+P_{2r})+\C(P_{2p})\label{eq: R2_3_bd_ag11}\\
R_1+R_2         &\le \C(P_1)+\C(P_{2p})\label{eq: R1pR2_5_bd_ag11}\\
R_1+R_2         &\le \C(P_{1z}+P_{2r})+\C(P_{2p}+a^2 P_{1c})\label{eq: R1pR2_3_bd_ag11}\\
R_1+R_2         &\le C_{12}+\C(P_{1c}+P_{2r})+\C(P_{2p})\label{eq: R1pR2_4_bd_ag11}\\
R_1+R_2         &\le \min\{C_{12},C(P_{1z})\}+\C(P_{2p}+a^2(P_{1c}+P_{2r}))\label{eq: R1pR2_1_bd_ag11}\\
R_1+R_2         &\le \min\{C_{12},C(P_{1z})\}+\C(P_{2p}+a^2 P_{1c})+C_{21}\label{eq: R1pR2_2_bd_ag11}\\
2R_1+R_2        &\le \min\{C_{12},C(P_{1z})\}+\C(P_{2p}+a^2 P_{1c})+C_{12}+\C(P_{1c}+P_{2r})\label{eq: 2R1pR2_1_bd_ag11}\\
2R_1+R_2        &\le \min\{C_{12},C(P_{1z})\}+\C(P_{2p}+a^2 P_{1c})+\C(P_1)\label{eq: 2R1pR2_2_bd_ag11}\\
R_1+2R_2        &\le \C(P_{1z}+P_{2r})+2\C(P_{2p})+\C(P_{1c}+P_{2r})\label{eq: R1p2R2_1_bd_ag11}\\
R_1+2R_2        &\le \C(P_{1z}+P_{2r})+\C(P_{2p})+\C(P_{2p}+a^2(P_{1c}+P_{2r})).\label{eq: R1p2R2_2_bd_ag11}
\end{align}
Again, the employed power allocation can help us to simplify the achievable region to the region described by (\ref{eq: R1_1_bd_ag11f})-(\ref{eq: R1pR2_3_bd_ag11f}).
This is because
\begin{align*}
(\ref{eq: R2_3_bd_ag11})&\geq (\ref{eq: R1pR2_5_bd_ag11})-0.5\\
(\ref{eq: R1pR2_3_bd_ag11})&= (\ref{eq: R1_1_bd_ag11})+(\ref{eq: R2_1_bd_ag11})\\
(\ref{eq: R1pR2_4_bd_ag11})&\geq (\ref{eq: R1pR2_5_bd_ag11})-0.5\\
(\ref{eq: R1pR2_2_bd_ag11})&\geq (\ref{eq: R1pR2_1_bd_ag11})-0.5\\
(\ref{eq: 2R1pR2_1_bd_ag11})&\geq (\ref{eq: R1_2_bd_ag11})+(\ref{eq: R1pR2_1_bd_ag11})-0.5\\
(\ref{eq: 2R1pR2_2_bd_ag11})&\geq (\ref{eq: R1_1_bd_ag11})+(\ref{eq: R1pR2_1_bd_ag11})-0.5\\
(\ref{eq: R1p2R2_1_bd_ag11})&\geq (\ref{eq: R2_3_bd_ag11})+(\ref{eq: R1pR2_5_bd_ag11})-0.5\\
(\ref{eq: R1p2R2_2_bd_ag11})&\geq (\ref{eq: R2_1_bd_ag11})+(\ref{eq: R1pR2_5_bd_ag11})-0.5.
\end{align*}
To obtain the preceding inequalities, we use the fact that $P_2+1\le a^2P_1$.

The achievable rate is compared below with the upper bound to prove ${\cal R}(1,\C(\frac{13}{3}))$ is achievable:
\begin{align*}
(\ref{eq: Ru1_1})-(\ref{eq: R1_1_bd_ag11f})&\leq 0.5\\
(\ref{eq: Ru1_1})-(\ref{eq: R1_2_bd_ag11f})&\le 1\\
(\ref{eq: Ru2_1})-(\ref{eq: R2_1_bd_ag11f})&\leq 0.5\\
(\ref{eq: Ru2_2})-(\ref{eq: R2_2_bd_ag11f})&\leq \C(\frac{13}{3})\\
(\ref{eq: Ru1pRu2_1})-(\ref{eq: R1pR2_1_bd_ag11f})&\leq 1.5\\
(\ref{eq: Ru1pRu2_2})-(\ref{eq: R1pR2_2_bd_ag11f})&\stackrel{(\star)}{\le} \C(9)\\
(\ref{eq: Ru1pRu2_2})-(\ref{eq: R1pR2_3_bd_ag11f})&\stackrel{(\diamond)}{\le} \C(11).
\end{align*}
To achieve $(\star)$, and $(\diamond)$, the upper bound (\ref{eq: Ru1pRu2_2}) is enlarged to $\C(2P_1P_2+2P_1+2P_2+a^2P_1)$, and $\C(3P_1P_2+3P_1+2P_2)$, respectively, as a consequence of $a^2\le P_2+1$. It is seen that the sum-capacity is determined up to 2 bits in this scenario since $(\ref{eq: R1pR2_1_bd_ag11f}) \le (\ref{eq: R1_2_bd_ag11f})+(\ref{eq: R2_2_bd_ag11f})$.

\end{document}